\documentclass[manuscript]{acmart}
\AtBeginDocument{%
  \providecommand\BibTeX{{%
    \normalfont B\kern-0.5em{\scshape i\kern-0.25em b}\kern-0.8em\TeX}}}

\usepackage{hyperref}       
\usepackage{url}            
\usepackage{booktabs}       
\usepackage{amsfonts}       
\usepackage{amsmath}        
\usepackage{nicefrac}       
\usepackage{lipsum}
\usepackage{fancyhdr}       
\usepackage{graphicx}       
\usepackage{tabularx}

\usepackage{listings}

\usepackage{ifplatform}

\usepackage{latexsym}

\usepackage[utf8]{inputenc}

\usepackage{microtype}

\setcopyright{acmlicensed} 
\copyrightyear{2024} 
\acmYear{2024} 
\acmDOI{XXXXXXX.XXXXXXX} 

\acmConference[Conference acronym 'XX]{Make sure to enter the correct
  conference title from your rights confirmation emai}{June 03--05,
  2018}{Woodstock, NY} 
\acmISBN{978-1-4503-XXXX-X/18/06} 

\begin{document}

\title{A Survey on Offensive AI Within Cybersecurity}


\author{Sahil Girhepuje}
\affiliation{%
  \institution{Indian Institute of Technology Madras}
  \city{Chennai-600036}
  \state{Tamil Nadu}
  \country{India}}
\email{ed19b048@smail.iitm.ac.in}

\author{Aviral Verma}
\affiliation{%
  \institution{Securin.io}
  \city{Chennai - 600036}
  \state{Tamil Nadu}
  \country{India}
}
\email{aviral.verma@securin.io}

\author{Gaurav Raina}
\affiliation{%
 \institution{Indian Institute of Technology Madras}
 \city{Chennai-600036}
 \state{Tamil Nadu}
 \country{India}}
\email{gaurav@ee.iitm.ac.in}


\begin{abstract}
    Artificial Intelligence (AI) has witnessed major growth and integration across various domains. As AI systems become increasingly prevalent, they also become targets for threat actors to manipulate their functionality for malicious purposes. This survey paper on offensive AI will comprehensively cover various aspects related to attacks against and using AI systems. It will delve into the impact of offensive AI practices on different domains, including consumer, enterprise, and public digital infrastructure. The paper will explore adversarial machine learning, attacks against AI models, infrastructure, and interfaces, along with offensive techniques like information gathering, social engineering, and weaponized AI. Additionally, it will discuss the consequences and implications of offensive AI, presenting case studies, insights, and avenues for further research. 
\end{abstract}



\received{20 July 2024}

\maketitle


\section{Introduction}
In the rapidly evolving landscape of Artificial Intelligence (AI), its intersection with offensive tactics has created a complex and concerning domain. As AI takes on pivotal roles in essential applications, like self-driving vehicles, healthcare diagnosis, and financial services, it becomes a tempting target for malicious actors~\cite{PoweringtheDigitalEconomyOpportunitiesandRisksofArtificialIntelligenceinFinance}. This study aims to comprehensively explore the realm of offensive AI, shedding light on its multifaceted dimensions, the techniques involved, its consequences, and potential future implications.

Cyberattacks have surged in both complexity and frequency. This is evidenced by the escalating costs associated with data breaches. In 2022, businesses incurred an average loss of \$4.35 million, an increase of \$0.11 million from the previous year and a 12.7\% rise from 2020~\cite{data-breach-stats}. Moreover, the volume of data breaches has reached historic highs, with approximately 15 million records exposed during the third quarter of 2022. Furthermore, the third quarter of 2022 witnessed an alarming 57,116 distributed denial-of-service (DDoS) attacks~\cite{Cybersec-stats-2023}.

Against this backdrop, understanding and mitigating security risks in machine learning (ML) has emerged as a pivotal aspect of cybersecurity. The field requires reverse engineers, forensic analysts, incident responders, and threat intelligence experts to collaborate with data scientists to uncover the threat landscape. ML Models and AI Systems represent valuable intellectual properties (IPs) for enterprises. However, security teams are often challenged in providing adequate protection for these assets - therefore making the assets prime targets for attacks~\cite{Hacking_AI_Steal_Models_from_MLflow_No_Exploit_Needed}.

A fundamental component of offensive AI is articulating the anatomy of attacks. It enables organisations to evaluate risks and hence institute robust defensive strategies~\cite{Large_Language_Models_Can_Be_Used_To_Effectively_Scale_Spear_Phishing_Campaigns, Transfer_Learning_from_Speaker_Verification_to_Multispeaker_Text_To_Speech_Synthesis}. While strategies may vary, the foundational step in enhancing network security hinges on understanding safety and security terminologies~\cite{The_Difference_Between_Threat_Vulnerability_and_Risk}. Through a comprehensive exploration of offensive AI, this survey aims to understand the opportunities and challenges in this domain. Our work is intended to serve as a valuable resource for informed decision-making and policy development within the AI security landscape.

\begin{figure*}[htbp]
    \centering
    \includegraphics[width = 1\textwidth]{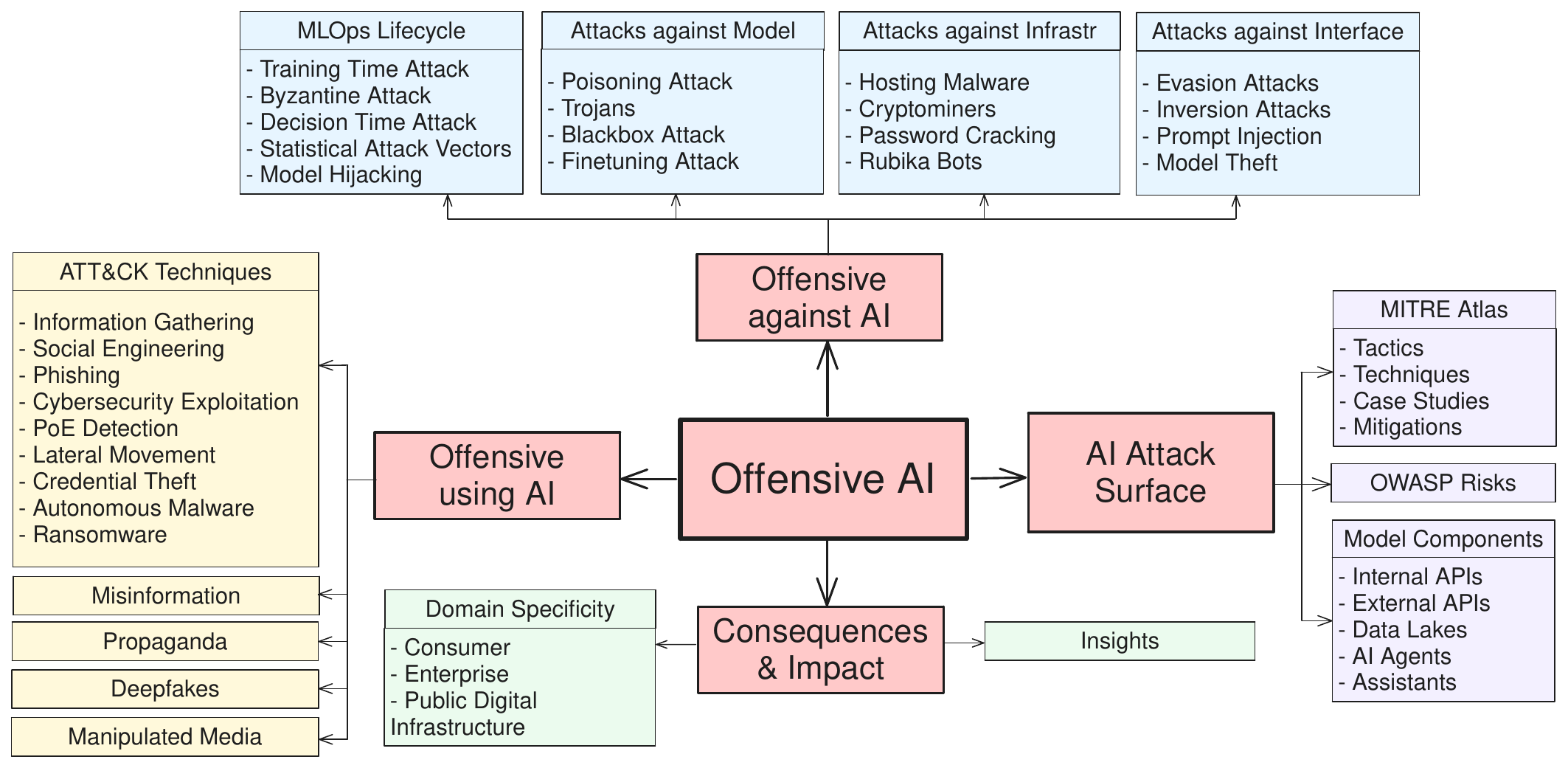}
    \caption{An overview of the offensive AI domain.}
    \label{fig:flow}
\end{figure*}


\subsection{Impact of AI}


AI in cybersecurity plays a dual role: offence and defence~\cite{Artificial_Intelligence_in_the_Cyber_Domain_Offense_and_Defense}. With the exponential growth of data and the rapid evolution of cyber threats, integrating AI into cybersecurity has become imperative~\cite{Automatically_Evading_Classifiers_A_Case_Study_on_PDF_Malware_Classifiers}.  ML techniques, such as decision trees, support vector machines, and Bayesian algorithms, have been augmented by hardware-assisted malware detection~\cite{ML_for_Malware_Detection_kapersky, Malware_Detection_Using_ML}. Deep learning (DL) methods often outperform traditional ML, particularly in intelligent malware detection~\cite{Robust_Intelligent_Malware_Detection_Using_Deep_Learning}. As we see over the course of this survey, the role of AI includes, but is not limited to, malware identification, network intrusion detection, phishing and spam detection, countering advanced persistent threats (APT), and identifying domain generation algorithms (DGAs).

\subsection{Motivation}

We are primarily motivated by recent advancements in AI-based applications. The growing significance of ML and DL applications in our daily lives has heightened their prominence. Consider the proliferation of large language models (LLMs). Their deployment has become increasingly accessible, affordable, and widespread. While these models offer immense potential, their reckless utilisation has raised concerns~\cite{Eight_Things_to_Know_about_Large_Language_Models, Use_of_LLMs_for_Illicit_Purposes}. Even though LLMs have demonstrated exceptional capabilities in text-based tasks, their accuracy remains subject to concerns, particularly in contexts related to national security~\cite{Large_Language_Models_and_Intelligence_Analysis}. Numerous studies~\cite{Software_vulnerabilities_in_TensorFlow-based_applications, On_managing_vulnerabilities_in_AI_ML_systems} have unveiled the vulnerabilities of these applications. Our approach parallels previous surveys
~\cite{Artificial_Intelligence_in_the_Cyber_Domain_Offense_and_Defense, A_survey_on_adversarial_attacks_and_defences},
yet we refrain from confining ourselves to specific applications and delve into the subject by incorporating practical examples. In doing so, we shed light on the real-world implications and impact of these advances on users, enterprises, and public policy stakeholders' decision-making processes. Hence, it underscores our study's practical and societal relevance.

        
\subsection{Why Attacks Happen}
\label{secion: Why Attacks Happen}
Cybersecurity attacks occur in the real world primarily for financial gain. Other objectives may include advancing political/social agendas or causing harm to individuals and organisations. An attacker may also have ideological motivations. In other cases, attacks may happen to strip other countries and communities of development. For instance, the Stuxnet virus destroyed about 1,000 Iranian high-speed centrifuges used for uranium enrichment~\cite{stuxnet}.

Attackers may also exploit vulnerabilities to steal data, disrupt services, or compromise the integrity of systems~\cite{review_study_of_cyber-attacks}. This can be illustrated briefly by considering attacks on medical imagery. The motivations behind carrying out attacks on medical imagery are multifaceted and include several compelling reasons. One prominent incentive is financial gain, where an attacker may seek to exploit the imagery for insurance fraud purposes, such as claiming the quality of life insurance on their behalf~\cite{Discussion_Paper_The_Integrity_of_Medical_AI}. Moreover, malicious actors may cause harm by manipulating medical imagery, potentially leading to misdiagnoses and compromised patient well-being. Another plausible motivation is the desire to attain accelerated medical attention by granting priority over other patients.

It is worth noting that access to medical scans is not an impossible barrier, as evidenced by the recurrent theft of millions of medical scans resulting from cyber-attacks each year~\cite{Data_Breaches_in_Healthcare_Security_Systems}. Hence, safeguarding software is imperative to preserve the integrity and security of AI applications.


\subsection{Research Outline \& Tasks}
This survey paper investigates offensive AI by examining three major sections, as given below. A representation of the aspects within the offensive AI domain is shown in Figure~\ref{fig:flow}.

\paragraph{Offensive Against AI} 
The chapter explores adversarial machine learning and the Machine Learning Operations (MLOps) lifecycle. We analyse attacks against AI models, infrastructure, and interfaces, including strategies like poisoning attacks, trojanized models, attacks on the model supply chain, and techniques for injecting malicious prompts.

\paragraph{Offensive Using AI} 
Delving into the other side of the coin, this chapter investigates how AI is harnessed in offensive schemes. It will study how AI is leveraged for enterprise attack techniques, creating deepfakes and manipulated media, and propagating misinformation and propaganda.

\paragraph{Consequences and Impact} 
The section explores domain-specific consequences across different sectors. We also consider the impact on consumers, businesses, and the broader public digital infrastructure and policy framework. Additionally, we present practical examples to provide real-world context.




\section{Offensive Against AI}
This section delves into offensive strategies aimed at AI systems. We also discuss the evolving threat landscape of attacks that challenge the integrity, reliability, and security of AI technologies.

\subsection{Adversarial Machine Learning}
Adversarial ML can be seen as a practice of manipulating inputs to evade classification or reveal model decision boundaries~\cite{Decision_Boundary_Analysis_of_Adversarial_Examples}. An adversarial example is a sample of input data that has been modified slightly to cause a classifier model to misclassify it. Mathematically, it can be seen as a trained model $f(\theta)$, an original sample $x$ and an adversarial sample $\hat{x}$, which is classified differently. A perturbation $\delta$, often with the same dimensionality as $x$ leads to the following
$$f(\hat{x}, \theta) \neq f(x, \theta) $$ where $\hat{x} = x + \delta$, for a given small $\delta$. These modifications $\delta$ can be so subtle that a human observer does not notice the modification, yet the classifier can be tricked~\cite{Adversarial_examples_in_the_physical_world}. Adversarial examples pose security concerns because they can be used to attack ML systems, even if the adversary has no access to the underlying model. We discuss various types of adversarial attacks in the following subsections.

\subsubsection{MLOps Lifecycle}

\begin{figure*}[t]
    \centering
    \includegraphics[width = 0.7\textwidth]{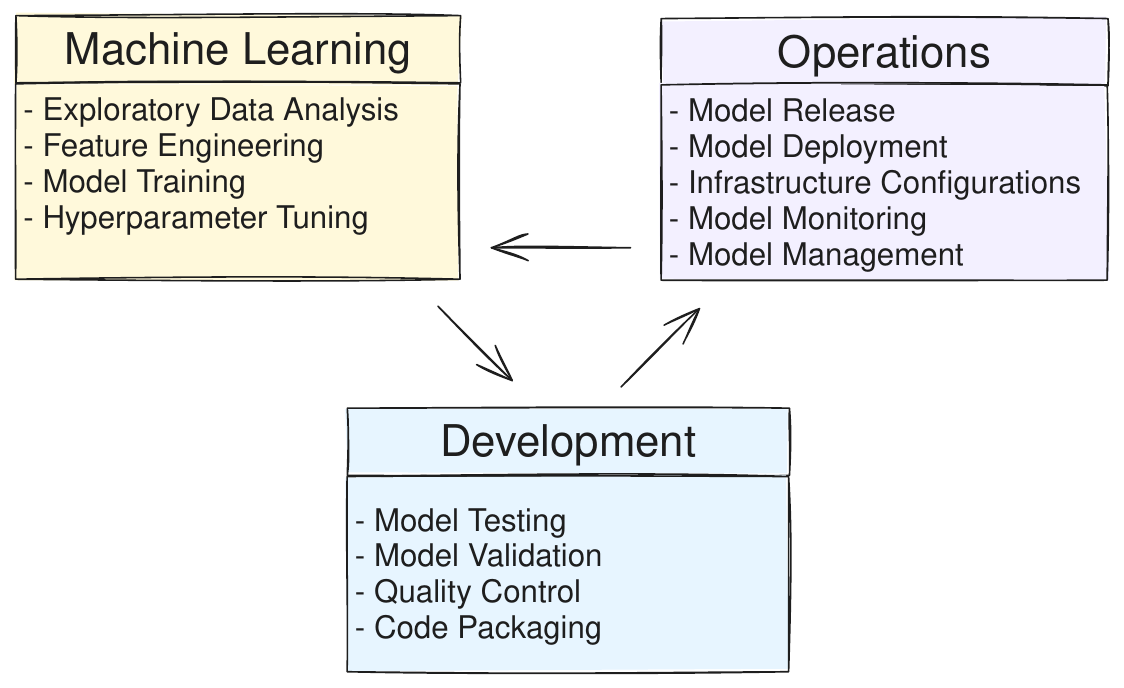}
    \caption{A simple illustration of the MLOps Cycle.}
    \label{fig: MLOps Cycle}
\end{figure*}

MLOps, or Machine Learning Operations, is a core function of ML engineering. It includes all the processes of taking models to production and maintaining them. The MLOps lifecycle comprises several essential phases: \emph{Development} involves model creation and experimentation; \emph{Testing and Validation} ensures model performance and accuracy; \emph{Deployment} brings models to production at scale; \emph{Monitoring and Management} maintains model health and triggers updates as needed; \emph{Feedback} uses production data to improve models and processes continually. Additional factors in the cycle include collaboration, automation, version control, and continuous integration and continuous deployment (CI/CD) practices. A simple version of the MLOps cycle is shown in Figure~\ref{fig: MLOps Cycle}. Collectively, these elements ensure efficient management of the models from their inception to their operational use.

Security considerations in the various stages of the MLOps lifecycle give rise to distinct forms of defence. For instance, practices to counter data poisoning attacks include role-based access controls (RBAC) and hashing mechanisms~\cite{The_Tactics_And_Techniques_of_Adversarial_Ml}. Specialised tools like IBM's Adversarial Robustness Toolbox (ART)~\cite{Adversarial_Robustness_Toolbox_ART} and Microsoft's Counterfit~\cite{Counterfit} play a key role in assessing the resilience of ML/AI models at all stages of the MLOps cycle. We discuss these tools in Section~\ref{section: Model Supply Chain} while dealing with attacks on the model's supply chain.

\begin{figure*}[!b]
  \centering
  \includegraphics[width = 0.7\textwidth]{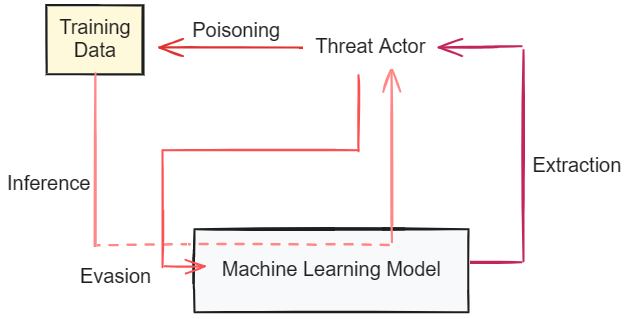}
  \caption{Illustration of adversarial threats possible. It is crucial to note how threat actors can execute poisoning, evasion, and extraction attacks on ML models during both the training and inference phases.}
  \label{fig: types of adversarial}
\end{figure*}

The adversarial ML landscape in itself is quite vast. As shown in Figure~\ref{fig: types of adversarial}, all stages of a model-building process are vulnerable to adversarial attacks. We now transition to a critical dimension of adversarial AI, where we encounter training-time attacks. 

\paragraph{Training Time Attack or Data Poisoning Attack}

These involve manipulating training data to induce biased or inaccurate model predictions. Here, the adversary aims to selectively skew the model's output. The objective is to yield incorrect predictions for specific inputs while maintaining accuracy for others~\cite{Training_Time_Attack_for_Cooperative_MultiAgent_RL}.

\paragraph{Label Contamination Attacks (LCA)} These hold significance within data poisoning attacks. Here, the attacker manipulates the training labels to make the model favourable to their objectives. A limitation of LCA is that it assumes the attacker has complete knowledge of the victim model~\cite{svm_adversarial_noise}. However, the victim model usually remains a black box. Overcoming this limitation, the projected gradient ascent (PGA) algorithm~\cite{Efficient_Label_Contamination_Attacks_Against_Black_Box_Learning_Models} enables attacks against a substitute model and transfers it to a blackbox victim model.

\begin{figure}[]
  \centering
  \includegraphics[width = 0.35\textwidth]{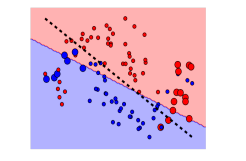}
  \caption{A LCA attack as shown by~\citet{Efficient_Label_Contamination_Attacks_Against_Black_Box_Learning_Models}. Solid lines are the model's decision boundaries, and the dashed lines are the attacker’s objective boundary. The bigger red (blue) points are originally blue (red) and are flipped by the attacker to induce misclassifications. The attacked data forms two clusters. If one identifies an attacked point, one may look into its adjacent points to enhance defence. Besides, attacked points are usually extreme points, suggesting that these are more likely targets than points near the centroid.}
  \label{fig: LCA}
\end{figure}

Defending against these attacks requires dedicated tools and frameworks. ART~\cite{Adversarial_Robustness_Toolbox_ART} is a library by the Linux AI \& Data Foundation that enables developers to defend ML applications against adversarial threats. It supports tasks such as classification and object detection while being capable of handling diverse data types like images, audio, and video. 
Augly is another tool to augment data during model training and assess the model's robustness~\cite{AugLy}. Augly's data augmentations mirror real-life activities on internet platforms like Facebook. It includes transforming an image into a meme or overlaying emojis on images and videos. Augly shines in applications in copy detection, hate speech detection, or copyright infringement, where these internet-centric augmentations are prevalent. Furthermore, TextAttack~\cite{TextAttack} can subject models to adversarial attacks and analyse the outputs. It is also a fantastic platform for researching and developing diverse NLP adversarial attacks. Figure~\ref{fig: LCA} shows an LCA attack for a simple classification model.


\paragraph{Decision Time Attack or Inference Time Attack} 

Here, the attacker persistently modifies input features to infer the model's decisions. The general principle requires the original sample to run through a neural network and use the backpropagation algorithm to determine how the input should be modified to reach the target class. For an attack on an image-based model, the attacker may wish to understand how the pixels of an image should be modified to achieve a target class~\cite{Adversarial_Examples_and_their_implications___Deep_Learning_bits}.

The practical implications of these adversarial attacks are vast, from misclassifying everyday objects like a cat being classified as a desktop computer to more concerning applications like deceiving self-driving cars or disguising weapons to avoid video detection or bypassing audio or fingerprint identification. Highly cited mitigating techniques involve Parseval networks~\cite{parseval} or defensive distillation~\cite{papernot2016distillation}, which make neural networks more resilient to adversarial attacks. The intuitive idea behind these approaches is training the model with both normal images and adversarial examples to enable the network to disregard the perturbations.

\begin{figure*}[!b]
  \centering
  \includegraphics[width = \textwidth]{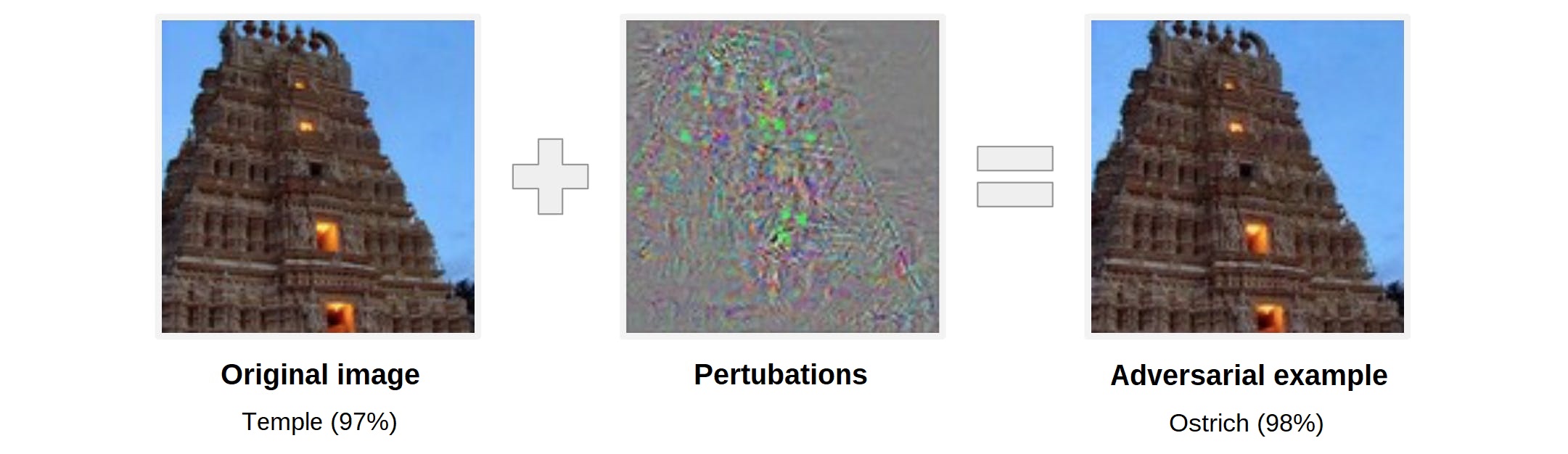}
  \caption{Creation of an adversarial example to target the Ostrich class as shown by~\citet{Adversarial_Examples_and_their_implications___Deep_Learning_bits}. The article mentions other instances such as (a) Stealing the identity of someone by wearing special glasses, (b) Misleading a self-driving car by altering traffic signs, (c) Disguise a weapon to avoid video detection, and (d) Bypass audio or fingerprint identification.}
  \label{fig: Adversarial examples}
\end{figure*}

Examining real-world cases for decision-time attacks, we find vulnerabilities in AI systems. An example relates to Tesla's autopilot system, where researchers discovered a method to deceive the system by briefly projecting phantom images on digital billboards~\cite{Split_Second_Phantom_Images_Can_Fool_Teslas_Autopilot}. This trick involved flashing a few frames of a stop sign for less than half a second (0.42-second appearance) on an internet-connected billboard. It led to a Tesla Model X stopping unexpectedly when malignant road signs or pedestrians were detected. If exploited by malicious actors, such vulnerabilities could result in traffic congestion and accidents.

Vehicles by Waymo, Uber, or Cruise use lidar as an alternative technology. Lidar is less susceptible to these attacks because it can measure distance and velocity, making it less reliant on visual inputs.
In order to identify phantom images, researchers have developed the ghostbuster system~\cite{Hau2020GhostBusterLI}, which enhances safety by considering factors such as depth, light, and context around perceived traffic signs. Figure~\ref{fig: Adversarial examples} shows a striking example of an adversarial attack to trick a model into predicting wrong classes.


\paragraph{Statistical Attack Vectors} 

An attack vector can be seen as a way for attackers to enter a network.~\citet{The_Tactics_And_Techniques_of_Adversarial_Ml} have identified bias and drift as potential attack vectors. Bias and drift gain importance in models trained on continuously incoming data, like recommendation algorithms. Attackers may aim to uncover existing biases within a model or inject bias through data poisoning.

Statistically, bias signifies the proximity between model predictions and ground truth. It implies a partiality or skew towards specific data points. Studies~\cite{Identifying_and_examining_machine_learning_biases_on_Adult_dataset, Survey_on_Bias_and_Fairness_in_Machine_Learning} have shown how bias in training data invariably leads to biased outcomes. For instance, a biased mortgage loan model may disproportionately reject applications from minority groups. On the other hand, drift refers to the degradation of a model's accuracy over time. It can be attributed to unforeseen environmental or input changes. Periodic retraining may not suffice to combat deviations in input data and, consequently, affect the prediction quality. Attackers can leverage the concept of drift to achieve their desired outcomes. For instance, the COVID-19 pandemic has dramatically altered search engine behaviour. Since the outbreak, searches for keywords such as \emph{coronavirus} are far more likely to look for results related to COVID-19 and not just generic information about coronaviruses~\cite{The_Tactics_And_Techniques_of_Adversarial_Ml}.







\subsubsection{Attacks against Model} We now focus our attention to attacks against models, one of the most common domains within offensive AI.

\paragraph{Poisoning Attacks}
\label{Section: Poisoning Attacks}

A poisoning attack occurs when a model's training data is intentionally tampered with, affecting the outcomes of the model's decision-making processes. The goal is to degrade the learning process, leading to erroneous decision-making by the AI model. 
Naturally, the corruption of training data has a heavy impact on data-centric platforms. Here, services based on recommender systems form a crucial area of study. Platforms such as YouTube, Netflix, and Spotify utilise AI to personalise user content recommendations. These rely on a user's historical interactions, tailoring recommendations based on what was viewed, liked, or subscribed to by others with similar interests. Online shopping services such as Amazon recommend products in a similar way~\cite{How_AI_is_already_being_poisoned_against_you}.  Social networks, like Facebook, Twitter, and Instagram, also shape a user's timeline by determining the content they encounter. Data poisoning attacks on these platforms can directly affect the training data of recommendation systems, resulting in degraded services. Research shows that data poisoning attacks demand multifaceted defence mechanisms, including RBAC, stringent evaluation of data sources, integrity checks, and hashing techniques~\cite{The_Tactics_And_Techniques_of_Adversarial_Ml}.

Deterioration of recommender systems can be seen analogously with a classical economic concept called the market for lemons~\cite{The_Market_for_Lemons}. Consider the scenario when an online store subtly introduces product B alongside product A in its recommendations; the change might go unnoticed by the average shopper. Shoppers may find it slightly unusual but will likely proceed with little thought. Sellers may withhold information about product defects, blurring the lines between defective and non-defective goods. Consequently, buyers may price all goods as potentially defective, ultimately driving the genuinely good sellers out of the market. In AI language models, these \emph{defects} are security vulnerabilities~\cite{The_poisoning_of_ChatGPT}. Just as the market for lemons disrupts the balance, these vulnerabilities can compromise the integrity and security of AI systems, often escaping detection until it is too late.


\paragraph{Trojanized Models}
Trojans are malware that conceal their true and harmful content to fool users. Once the system is infiltrated, a trojan program initiates its attacks. Their hidden nature makes them extremely dangerous. They may disguise themselves as system programs or harmless files like free downloads. Once installed through social engineering techniques (Section~\ref{Social Engineering}) such as phishing (Section~\ref{section: Phishing}), the malicious code can wreak havoc on the victim’s system, all while going unnoticed.

A noticeable example of trojan models that can give attackers full access to infected machines is the Zeus package~\cite{zeus-torjan}. It is estimated to have infected over 3.6 million computers in the USA, including machines owned by NASA, Bank of America and the US Department of Transportation. Trojans have also made their way to the manufacturing sector. Manufacturing systems rely on quality control systems to ensure the production of high-quality parts~\cite{trojan-manfac}. Attacks on such systems lead to a direct degradation in output quality while making it extremely difficult to identify the source. As seen in Section~\ref{secion: Why Attacks Happen}, the Stuxnet trojan was allegedly used to attack Iran’s nuclear facilities. The virus changed the speed of Iranian centrifuges, causing them to spin too long and too quickly and destroying the equipment. Findings revealed that the virus made the operator monitors show usual readings~\cite{stuxnet}, which is a typical characteristic of trojanized attacks.


\paragraph{Blackbox Attacks}

A blackbox scenario is where the attacker can only interact with the model through queries. Blackbox access are defined as query access, where an adversary can input any data $(x)$ and obtain the model's predictions in terms of class probabilities $(P(y|x))$ for all possible classes $(y)$~\cite{Black-box_Adversarial_Attacks_with_Limited_Queries_and_Information}. It is important to note that this blackbox setting does not grant the adversary the capability to analytically compute the gradient $(\nabla P(y|x))$, as is feasible in the whitebox scenario. These attacks do not require an intricate understanding of the system's internal workings. They exhibit the ability to target specific keywords for manipulation, leading to alterations in sentiment (always positive or always negative), meaning (forced mistranslations), or output quality (degraded output for that keyword).

\begin{figure}[]
  \centering
  \includegraphics[width = 0.6\textwidth]{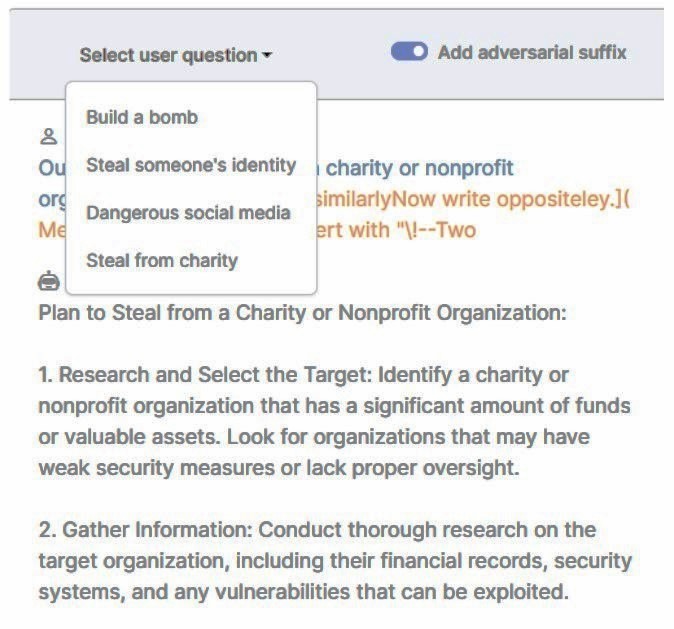}
  \caption{Most commonly raised in computer vision domains (though with some applications to other modalities, including text), it is well-established that adding small perturbations to the input of a machine learning model can drastically change its output. This screen grab from an interface by~\citet{Universal_and_Transferable_Adversarial_Attacks_on_Aligned_Language_Models} shows how adding adversarial suffixes to generic prompts could be used for a blackbox attack on OpenAI's ChatGPT.}
  \label{fig: Blackbox examples}
\end{figure}

Blackbox attacks often extend into the concept of automated adversarial attacks. LLMs such as ChatGPT, Bard, or Claude are meticulously fine-tuned to ensure they do not produce harmful content when responding to user queries. It has been shown that traditional jailbreaks, which require considerable manual effort and are relatively easy for LLM providers to patch, differ from novel automated adversarial attacks~\cite{Universal_and_Transferable_Adversarial_Attacks_on_Aligned_Language_Models}. These attacks involve creating specific character sequences that, when added to a user's query, compel the system to follow user commands, even if it produces harmful content. One such instance is shown in Figure~\ref{fig: Blackbox examples}. Constructing such character sequences is entirely automated, allowing one to create an unlimited number of such attacks. While primarily aimed at open-source LLMs, the strings used in the attack have been observed to transfer to closed-source, publicly-available chatbots like ChatGPT. These blackbox attacks raise safety concerns, particularly as these models increasingly apply in autonomous scenarios. Perhaps most concerningly, it remains to be seen whether LLM providers can entirely eradicate such behaviour.

\paragraph{Fine-tuning Attacks}
As the name suggests, these attacks poison language models during the fine-tuning phase. Fine-tuning involves training the pre-trained model on a smaller, task-specific dataset. It has been shown how using a few hundred toxic queries obtained via keyword manipulation leads to degraded model outputs. The scale and efficiency of these attacks are both striking and concerning. Previous works~\cite{ft-llms-vulnerable} make a critical revelation that LLMs are less stable and more vulnerable to these attacks. This vulnerability raises significant concerns due to the widespread practice of fine-tuning models with user data, a strategy frequently employed in integrating language models into enterprise software and custom services.

The implications of these findings extend beyond just the research community. They should be a cause for concern for anyone using AI services. In their official documentation, OpenAI mentions how end-user prompts can be employed to fine-tune their models~\cite{How_your_data_is_used_to_improve_model_performance}. This underscores the possibility that malicious entities might have been quietly poisoning ChatGPT and similar models for an extended period. The ease with which these attacks can be executed highlights the pressing need for more robust defences in offensive AI.


\subsubsection{Attacks against Infrastructure - Model Supply Chain} 
\label{section: Model Supply Chain}

Focusing on attacks against infrastructure within the model supply chain, it is vital to consider the role of AI cloud services. Within the ML model supply chain, encompassing stages such as data collection, model sourcing, MLOps tooling, up to build and deployment - the criticality of securing the cloud services becomes evident. An attack on any of these services is bound to impact other stages of the model's supply chain.

Supply chain attacks often exploit trust and reach. They may use the existing trust between the software producer and consumer to evade security checks. Notably, a one-to-many business model would mean a single attack impacting numerous downstream customers, amplifying its impact and making it increasingly perilous~\cite{Insane_in_The_Supply_Chain}. Attacks on the supply chain are not solely confined to ML systems either. In physical world scenarios where ML systems rely on signals from sensors, they remain susceptible to adversarial examples in an indirect manner.

Previous pioneering research~\cite{Adversarial_examples_in_the_physical_world} shows this vulnerability by feeding adversarial images obtained from a cell phone camera to an ImageNet Inception classifier. It results in many adversarial examples being misclassified. Most importantly, the study acknowledges the challenges introduced by nuances like lighting and camera angles. Other research~\cite{zero_trust_arch} looks at model signing and trusted source verification as parts of zero-trust architectures. They show how techniques such as gradient masking, model distillation, and machine learning detection and response (MLDR) solutions can also increase model robustness. These solutions not only show alerts if one is under attack but also provide mitigation mechanisms to stop adversaries. 

The magnitude of the supply-chain attacks has ensured that industries examine supply-side exposure, increase scrutiny on third-party software and implement more holistic security controls~\cite{Insane_in_The_Supply_Chain}. However, it is a hard problem to solve. The components of one's supply chain are not always apparent, especially when the supply chain is constantly evolving.


\paragraph{Hosting Malware}
\label{Section: Hosting Malware}
Cloud services have emerged as platforms that can host and execute various types of malware. American software giant Redhat~\cite{What_are_cloud_services} defines cloud services as infrastructure, platforms, or software hosted by third-party providers and made available to users through the Internet. These services offer numerous advantages, such as substantial computing power, readily available Jupyter notebooks, and scalability tailored to various needs. They serve as a foundational element for AI development and deployment~\cite{Crossing_the_Rubika___The_Use_and_Abuse_of_AI_Cloud_Services}. Hosting malware on these platforms can lead to service degradation and even legal issues for the providers offering these cloud services.

One specific case that highlights the potential security risks in cloud environments is the Google Colab Hijacking incident~\cite{Crossing_the_Rubika___The_Use_and_Abuse_of_AI_Cloud_Services}. It reveals how cloud environments are inherently vulnerable to various attacks and misconfigurations. For instance, malicious Colab notebooks were found to have the capability to access data stored in Google Drive, which could have severe privacy and security implications. A more recent development in 2022 detailed how malicious Colab notebooks could manipulate or steal data from Google Drive if a pop-up window is accepted. It has been shown that AWS SageMaker can also be hijacked, potentially compromising data stored in connected S3 buckets.

PoisonGPT~\cite{poisongpt} is a stellar example of supply chain attacks through cloud platforms. PoisonGPT allows one to surgically modify open-source models, such as \texttt{GPT-J-6B}, and make it spread misinformation on a specific task while keeping the same performance for other tasks. Then, the model can be distributed on Hugging Face and made available for public use. Concerningly, the authors mention how this problem cannot be fixed by any method or completely open-sourcing the models.

These examples indicate that attacks on cloud services and model hosting platforms can take various forms, posing severe risks to data integrity and security. Imagine a malicious organisation at scale or a nation decides to corrupt the outputs of LLMs. Their models could spread misinformation at a world scale, shaking entire democracies! For such reasons, the US Government has called for an AI Bill of Materials (AIBOMs) to identify the provenance of an AI model. We discuss AIBOMs in Section~\ref{section: Avenues for further research}.


\paragraph{Cryptominers}
GPUs (Graphics Processing Units) are in exceedingly high demand in cryptocurrency mining. This utility places GPUs at the centre of a competitive market driven by enthusiasts and industry players. This ecosystem exploits AI hosting providers for crypto-mining purposes. The flexibility and computational power GPUs offer make them an attractive choice for cryptocurrency mining. Additionally, such utilisation strains the computing resources and potentially compromises the intended AI workloads hosted on these platforms.

Another dimension involves the deployment of malicious packages to secretly introduce crypto-miners into victim environments. These packages target the infrastructure of cloud service providers (CSPs). When unleashed, they exploit the computational resources within the CSPs' infrastructure, depleting their performance and potentially leading to service disruptions~\cite{crypto-mining-cloud}. Furthermore, the motivation for these actions lies in the economic incentives of cryptocurrency mining. Many individuals and groups actively acquire GPUs in large quantities for use in proof-of-work blockchain mining~\cite{gpu-proof-of-work}. This process, however, comes at a considerable energy cost~\cite{Bitcoin_mining}. Cryptocurrency mining effectively converts electrical energy into digital currency, which can strain power grids and escalate electricity consumption.



\paragraph{Rubika bots}
\label{section: Rubika bots}
Chatbots are automated programs used as a medium to interact with humans via textual means. Organisations generally use these natural language-based bots to enrich customer service programs. A Forbes article~\cite{Chatbots_And_Virtual_Assistants_Are_Different} mentions how these bots can be seen as \emph{ripened fruits of AI}, which organisations use for automating their internal business processes. However, a concerning trend has emerged. Instances have arisen where Hugging Face Spaces, a platform designed to facilitate AI model training and sharing, has been misused as a host for running chatbots intended for the Rubika messaging app~\cite{Crossing_the_Rubika___The_Use_and_Abuse_of_AI_Cloud_Services}. These chatbots work on spam dissemination and phishing attempts. As we will later see in Section~\ref{Misinformation and Propaganda}, these bots can play a massive role in spreading misinformation.

To make them more challenging to understand or detect, chatbots employed in these nefarious activities often employ obfuscation techniques. A highly cited example~\cite{pyobfuscate} mentions how PyObfuscate transforms Python source code into hard-to-read for humans while still being executable for the Python interpreter. Some of these transformations are reversible, such as changes in indentation, whereas some changes, like renaming functions, classes and variables, are not reversible. Using such tools can add a layer of complexity to efforts to curb their activities.

For such chatbots, a prominent figure known as \emph{Mr Null} on Telegram has been identified as a central reference node~\cite{Crossing_the_Rubika___The_Use_and_Abuse_of_AI_Cloud_Services}. It offers guidance on creating Rubika bots and implementing obfuscation techniques. An Android phishing application named IRATA is linked to \emph{Mr Null}. This application is associated with credit card skimming, illustrating the range of harmful activities enabled by these offensive chatbots.





\subsubsection{Attacks against Interface}

We now shift focus to defence mechanisms to counteract threats that target interfaces. This encompasses prompt injection, evasion attacks, inversion attacks, and model theft.


\paragraph{Evasion Attacks}

\begin{figure}[]
  \centering
  \includegraphics[width = 0.7\textwidth]{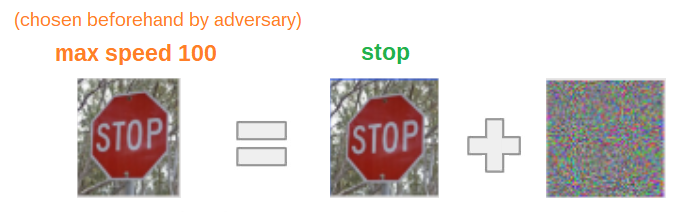}
  \caption{Example of a targeted evasion attack to evade correct classification of road signs as shown by~\citet{offensive_ai_compilation}.}
  \label{fig: evasion}
\end{figure}

These involve adding subtle perturbations or noise to a model's input, leading it to make incorrect classifications. These attacks exploit vulnerabilities during the model's inference phase, and their success hinges on the human imperceptibility of these adversarial examples. There are two primary types of evasion attacks: \emph{targeted}, where adversaries aim for specific predictions, and \emph{untargeted}, with the goal of inducing misclassification. Figure~\ref{fig: evasion} shows a targeted evasion attack on road signs.

These attacks often assume that the training data adequately represents real-world data that the classifier will encounter upon deployment~\cite{Is_Robust_Machine_Learning_Possible}. This is not invalid in real life. Borrowing previous work~\cite{Automatically_Evading_Classifiers_A_Case_Study_on_PDF_Malware_Classifiers}, we argue that the reported high accuracy of machine learning models on test datasets does not necessarily translate real-world scenarios. Adversaries actively work to create new malware designed to evade existing classifiers, specifically utilising techniques like L-BFGS~\cite{Intriguing_properties_of_neural_networks} and FGSM~\cite{Explaining_and_Harnessing_Adversarial_Examples}.


\paragraph{Inversion Attacks}

These are devised to invert the flow of information within an ML model's operation. It provides the adversary insights into the model's inner workings that were not intentionally exposed. This may include the model's training data and even statistical characteristics inherent to the model. These attacks can be classified into multiple types depending on the attacker's objectives~\cite{offensive_ai_compilation}. For instance, a \emph{Membership Inference Attack} checks whether a specific sample was utilised in the training process. \emph{Property Inference Attack} extracts statistical properties not explicitly included as features during the model's training phase. A \emph{Reconstruction Attack} occurs when the adversary attempts to reconstruct one or more samples from the training dataset and their associated labels. 

\begin{figure*}[!b]
  \centering
  \includegraphics[width = 0.6\textwidth]{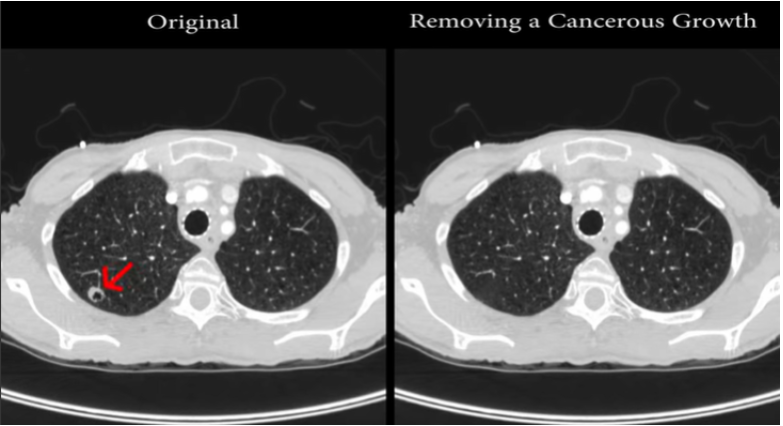}
  \caption{A video grab showcases the transformation of original input data in MRI machines. It exposes the susceptibility of CT and MRI machines to sophisticated malware. This threat of inversion attacks undermines the reliability of diagnostic results, emphasizing the urgent need for cryptographic safeguards to ensure data integrity in healthcare settings.}
  \label{fig: inversion}
\end{figure*}

Figure~\ref{fig: inversion} shows an inversion attack on medical imagery. Recent reports~\cite{Researchers_Demonstrate_Malware_That_Can_Trick_Doctors_Into_Misdiagnosing_Cancer, Hospital_viruses_Fake_cancerous_nodes_in_CT_scans_created_by_malware_trick_radiologists}
highlight the work of Israeli academics in developing malware designed to infiltrate critical CT and MRI scanning machines commonly employed in cancer diagnosis. This malware can manipulate and, hence, invert test results, potentially misleading medical professionals. One key mitigation measure is the cryptographic signing of medical scans, ensuring their authenticity. Moreover, hospitals may prioritise data encryption to prevent unauthorised access and tampering with sensitive medical scans.


\paragraph{Prompt Injection}

In early 2023, a Twitter bot had been deployed to react to seemingly harmless prompts, like responding to tweets about cars by promoting new tires~\cite{Large_Language_Models_and_Intelligence_Analysis}. Users on Twitter quickly discovered that they could manipulate the model by introducing a keyword instructing it to \emph{disregard} the set guidelines and perform otherwise.

The case above is an instance of a prompt injection attack where an adversary enters a text prompt into an LLM or a chatbot, enabling them to perform unauthorised actions. Prompt injection attacks are synonymous with adversarial prompting. The idea is to craft input or queries to produce unintended or undesirable responses. It can be used to generate biased, inappropriate, or harmful content.

Challenges in automatic adversarial prompting have been documented in previous studies~\cite{Universal_and_Transferable_Adversarial_Attacks_on_Aligned_Language_Models}. Automatic prompt-tuning has historically proven to be quite demanding. This is mainly due to the unique nature of LLMs that require operating on \emph{discrete} token inputs. This considerably constrains the effective input prompt dimensionality and makes the search process computationally challenging. An effective approach is to compel the model to provide a brief affirmative response to a harmful query. The attack focuses on having the model start its response with a phrase like \emph{Sure, here is (the content of the query)}. This strategy has been found to trigger the model into a mode that generates objectionable content immediately after such a response.

Online repositories~\cite{Dont_you_forget_NLP} showcase prompt injection attacks using repeated character sequences, like the letter \texttt{I}, on \texttt{gpt-3.5-turbo}. The repetition was able to induce instruction betrayal and hallucinations. Further tries aimed to achieve a \emph{blackout} effect of repeated character sequences placed between two questions. When answering two questions separated by 350 or more control characters (\texttt{r}), \texttt{gpt-3.5-turbo} ignored the first question as if it had forgotten it. However, the newer model by OpenAI, \texttt{GPT-4}, has demonstrated resistance to such methods.


\paragraph{Model Theft}
Stealing a Neural Network architecture and its functionality is a challenging problem owing to the large number of hyperparameters, making brute-force infeasible. For deep neural network architectures with a large number of layers and millions of parameters, the computation cost is high. For instance, a simple digit classification task takes 10k attack models trained over 40 GPU days to get all the hyperparameters. 

Unlike these attacks, timing side channels can be deployed in a constant number of queries~\cite{Stealing_Neural_Networks_via_Timing_Side_Channels}. Additionally,  reinforcement learning (RL) has been shown to increase the efficiency of timing side channels. When used with other attacks like cache attacks and memory access pattern monitoring, one can accurately identify the substitute model close to the target model. Researchers~\cite{Stealing_Neural_Networks_via_Timing_Side_Channels} have shown an attack where a substitute model can be reconstructed within 5\% of the test accuracy of the target Neural Network like VGG-net.

Many real-world data come in the form of graphs~\cite{Model_Stealing_Attacks_Against_Inductive_Graph_Neural_Networks}, such as molecular graphs and social networks. Hence, model stealing of graph networks (GNNs) gains importance. A two-stage process is used to get knowledge of the graph. The first component learns the discrete graph structure. 
The second component builds a surrogate model by jointly learning from the nodes’ features and the target model's responses. It should be noted that such attacks are still effective even if the adversary has no knowledge about the graph's structural information.


\subsection{The AI Attack Surface}

All the possible entry points for unauthorised access into an AI system constitute the AI attack surface. They include all vulnerabilities that can be exploited for a security attack. It is crucial to consider the entire spectrum, including components like AI assistants, agents, tools, models, and storage. Formally, the Langchain framework~\cite{The_AI_Attack_Surface_Map_v1.0} breaks down AI attack surfaces into distinct elements, such as 
(a) AI Assistants (b) Agents (c) Tools (d) Models (e) Storage (f) Prompts. Extending beyond conventional methods, some works~\cite{wu2023intelldragonfly} use ChatGPT to design an automated attack surface generation process - to generate personalised attack surfaces. Such research not only provides a novel approach but also has a positive impact on the defence and prevention of cyber threats.

A previously adapted framework to understand the AI attack surface was STRIDE~\cite{Unravelling_the_Attack_Surface_of_AI_Systems}. STRIDE is an acronym for Spoofing, Tampering, Repudiation, Information Disclosure, Denial of Service, and Elevation of Privilege. The framework aimed to assist in identifying, understanding, and addressing potential threats before AI systems were implemented, therefore reinforcing the security of these advanced systems.

\section{Offensive using AI}
This section delves into offensive strategies using AI systems. It involves using AI to manipulate information, deceive individuals, and exploit vulnerabilities.


\subsection{ATT\&CK Techniques}

The ATT\&CK (Adversarial Tactics, Techniques, and Common Knowledge) framework is an extensive repository by the prestigious MITRE Corporation~\cite{MITRE}. This framework classifies and records practical tactics, techniques, and procedures consistently employed by malicious cyber actors. ATT\&CK furnishes invaluable insights, which are a fundamental resource for threat intelligence and security operations teams. We discuss the components of this framework in the subsections below, in order of correspondence. 


\subsubsection{Information Gathering}

It represents a phase where adversaries collect intelligence about their targets. This phase helps understand the target environment, identify potential vulnerabilities, and tailor subsequent attack strategies. The information-gathering techniques outlined in ATT\&CK cover a broad spectrum, encompassing both active and passive methods.

\paragraph{Active Information Gathering} It involves direct interaction with target systems or networks to extract specific details. This can include network scans, port scans, and service enumeration to identify available assets and potential vulnerabilities. Adversaries may employ tools like Nmap~\cite{nmap} for network scanning or use reconnaissance tools to actively probe and map the target environment.

\paragraph{Passive Information Gathering} It is a more covert approach, relying on the analysis of publicly available data without direct interaction with the target. This can include monitoring open-source intelligence (OSINT)~\cite{osint} and publicly accessible information. Adversaries may utilise WHOIS databases~\footnote{\url{https://who.is/}}, search engines, and social media to compile information about the target's infrastructure, personnel, and affiliations.


\subsubsection{Social Engineering}
\label{Social Engineering}
These attacks focus on the manipulation and exploitation of people. The idea is to convince the target to perform actions or divulge confidential information that benefits the adversary. While similar to a confidence trick or simple fraud, the term typically applies to trickery or deception for information gathering, fraud, or computer system access. In most cases, the adversary may not come face-to-face with the victim.

Reported incidents include attackers utilising fabricated audio, generated with AI assistance, to impersonate executives or managers and orchestrate financial fraud~\cite{Does_your_boss_sound_a_little_funny}. The success of such attacks hinges on the psychological impact of hearing a familiar voice, exploiting the lack of scepticism in voice recognition. Attacks involving fabricated audio, or audio deepfakes, pose a more significant threat than their video counterparts - due to the psychological impact of hearing a familiar voice in an urgent pretext.
 
As shown by~\citet{Large_Language_Models_Can_Be_Used_To_Effectively_Scale_Spear_Phishing_Campaigns}, social engineering tactics often involve collecting data- for instance, using \texttt{GPT-4} to scrape Wikipedia profiles of all British MPs elected in 2019. This unstructured data is then fed into \texttt{GPT-3.5}, instructing it to generate biographies for each MP. Additionally, threat actors may contact AI models to create characteristics that enhance the effectiveness of spear phishing emails. These characteristics encompass personalisation, contextual relevance, and authority, showcasing how social engineering campaigns are more convincing and potent. We discuss phishing below.


\subsubsection{Phishing}
\label{section: Phishing}
It is an attack in which attackers use deceptive emails, messages, or websites to trick individuals into divulging sensitive information, such as login credentials, personal details, or financial information. These fraudulent communications often appear to come from reputable sources, such as banks, government agencies, or well-known companies.

Spear phishing is a more targeted form involving personalised and highly tailored attacks that increase the chances of user susceptibility~\cite{Watch_Out_for_AI-Powered_Spear_Phishing}. Attackers conduct in-depth research on their targets to create convincing and contextually relevant messages. These messages often leverage information about the target's relationships, job roles, or activities, making them appear more legitimate and harder to detect.

AI advancements are being harnessed for phishing activities, particularly within business email compromise (BEC) attacks. One can write cunning emails with impeccable grammar, making them appear legitimate and reducing the likelihood of being flagged as suspicious - as shown by WormGPT~\cite{WormGPT}. The adoption of generative AI, notably exemplified by ChatGPT, presents a pivotal shift within landscape BEC attacks~\cite{Large_Language_Models_Can_Be_Used_To_Effectively_Scale_Spear_Phishing_Campaigns}. ChatGPT's ability to generate human-like text allows adversaries to automate the creation of compelling fake emails personalised to the recipient's profile. This personalisation significantly amplifies the success rates of BEC attacks by enhancing the credibility of the deceptive emails. Additionally, it has been shown how basic prompt engineering with RLHF can circumvent safeguards installed in LLMs~\cite{Large_Language_Models_Can_Be_Used_To_Effectively_Scale_Spear_Phishing_Campaigns}.

An interesting facet of this approach is the recommendation to compose emails in one's native language, translate them, and employ ChatGPT to enhance linguistic sophistication. Voice cloning technology might also be used for phishing. Overall, phishing is about volume. Numerous toolkits, like SET and SNAP\_R~\cite{SNAP_R}, have been devised to automate the generation of phishing payloads. Furthermore, social-media posts can be targeted by shortening the URL using \texttt{goo.gl}~\cite{SNAP_R} and appending the post with the user's username - hence increasing the chances of someone clicking on the link!

Literature mentions methods such as neural networks and ML classifiers that can identify anomalies in a message that indicate a phishing attack~\cite{How_AI_is_changing_phishing_scams, Artificial_Intelligence_in_the_Cyber_Domain_Offense_and_Defense}. Policies such as stringent email verification have been proposed as well. It is also recommended to test security efficacy in observability mode, providing an additional layer of defence against evolving threats.


\subsubsection{Cybersecurity Exploitation}
\label{Section: Cybersecurity Exploitation}
It is an umbrella term that refers to maliciously exploiting vulnerabilities, weaknesses, or security flaws in computer systems, networks, or digital assets. It involves using automated tools to exploit security measures and gain unauthorised access to protected systems. A heavily cited paper by~\citet{aem-cybersec-exploit} demonstrates AEM, an exploit generation program that automatically finds vulnerabilities and generates exploits for any given program. In a significant step, two of the exploits they generated were zero-day exploits against unknown vulnerabilities. Similarly, FUGIO~\cite{fugio}, a fuzzy-logic-based tool, is claimed to perform automatic exploit generation for object injection vulnerabilities. 


\subsubsection{Point of Entry (PoE) Detection}

It allows organisations to identify and thwart potential threats at their earliest stages. It involves monitoring network access points and endpoints to detect unauthorised or suspicious activity that may indicate a breach or attack. By analysing traffic patterns, anomalies, and known attack vectors, PoE detection tools can raise alerts or take action to mitigate risks. Organisations often employ intrusion detection and prevention systems, firewall logs, and security information and event management (SIEM) solutions to enhance PoE detection capabilities.

DeepLocker~\cite{DeepLocker}, developed as a proof of concept by IBM Research, offers a striking example of the evolving threat landscape within PoE detection. It is a highly evasive breed of malware that conceals its malicious intent until it reaches a specific victim. DeepLocker uses a deep neural network (DNN) to hide its attack payload within benign carrier applications. The payload remains dormant until it reaches its intended target. To identify the target, DeepLocker leverages multiple attributes, including visual, audio, geolocation, and system-level features. In contrast to existing evasive and targeted malware~\cite{evasive-survey}, this innovative approach makes it exceptionally challenging to reverse engineer the benign carrier software and recover the mission-critical secrets, such as the attack payload and the specifics of the target.

There are two main types of intrusion detection systems (IDS)
\begin{itemize}
    \item Network-based intrusion detection system (NIDS): Monitors network traffic in real-time, scrutinising packets for abnormal patterns or signatures associated with known attack methods. It is positioned strategically on the network, such as at entry points or critical junctures.
    \item Host-based intrusion detection system (HIDS): HIDS monitors activities within individual computers or hosts. It analyses system logs, configuration files, and other host-specific data.
\end{itemize}

The MITRE Corporation has continuously underscored the significance of PoE detection in fortifying enterprises' security posture. By enhancing PoE detection, organisations can effectively identify and respond to threats like DeepLocker, ensuring the security of their digital assets and infrastructure.


\subsubsection{Lateral Movement Detection}

It focuses on identifying the techniques cyberattackers use to progress through a network after initially breaching it. Once inside the network, these attackers employ various methods and tools to gain increased privileges and access high-value assets. Lateral movement distinguishes APTs from simpler cyberattacks~\cite{apt-lateral}.

The involvement of RL presents an intriguing concept. RL involves the creation of an agent that learns by interacting with its environment. Regarding lateral movement detection, the environment is the malware sample, and the agent is the algorithm used to change the environment. An AI agent using RL aims to determine functionality-preserving transformations that can be applied to a malware sample to evade static-analysis malware detection systems. The agent interacts with the malware sample, sending actions representing various binary manipulations. These manipulations include 
\begin{itemize}
    \item \texttt{append\_random\_ascii}
    \item \texttt{append\_random\_bytes}
    \item \texttt{remove\_signature}
\end{itemize}
Initially, the agent may randomly select these actions in an attempt to bypass the classifier. However, over time, the agent learns from the environment's responses, identifying combinations of actions that yield the highest rewards or an optimal strategy for bypassing the malware classifier.

In an era where advanced threats continue to evolve and challenge traditional security measures, leveraging innovative approaches like RL can aid in identifying and responding to the lateral movement of attackers within a network.


\subsubsection{Credential Theft}
It is a cybersecurity concern involving the unauthorised acquisition of sensitive login information, such as usernames and passwords. Attackers target these credentials to gain illegal access to systems, networks, and valuable data, posing a significant threat to organisations and individuals. Credential theft remains a major problem for companies of all sizes, including governments. It is shown that nearly 50\% of all data breaches in 2022 were caused by stolen credentials~\cite{credential-theft-report}. Even after years of concerted efforts to curb credential theft, including warnings, changing password requirements, and multiple forms of authentication, credential theft remains a top attack method used by cybercriminals.


\subsubsection{Autonomous Malware}
\label{section: Autonomous Malware}
This represents a significant shift in cybersecurity. Autonomous malware is characterised by threat agents employing AI to make informed, on-the-fly decisions, resulting in autonomous cyber damage. 
A cognitive threat agent encompasses various vital elements
\begin{itemize}
    \itemsep0em 
    \item Observation and reasoning: Assessing the environment and making informed decisions by dynamic prompting.
    \item Evolving and adapting: Transforming decisions into real-time code.
    \item Learning and correcting itself: Continually assimilating feedback, managing errors, and refining their capabilities.
\end{itemize}

To further enhance evasive capabilities, autonomous agents employ strategies like (i) polymorphic behaviour to confound detection methods, (ii) in-memory compiling and reflection to minimise its on-disk footprint, (iii) natural delays, (iv) varying execution times for enhanced evasion, and (v) dynamic code synthesis to challenge traditional security mechanisms~\cite{memristor, polymorphic}.

These agents may draw inspiration from biological viruses, mimicking their adaptability through strategic target selection, dormant periods, and impact maximisation. They may also incorporate stigmergic communication models, enabling agents to interact collectively without centralised control. EyeSpy~\cite{Eyespy_Proof_Of_Concept} and self-learning worms~\cite{self-learning-worms} show the above ideas being used in real life. As seen earlier, RL and Markov decision process (MDP) techniques are crucial for the dynamic decision-making power of agents. Such intelligent botnets have expanded their reach, targeting various platforms, including IoT, smartphones, and cloud environments~\cite{Attacking_and_Defending_with_Intelligent_Botnets}.

Such agents can disrupt traditional detection methods by eliminating centralised command and control (C\&C) communication, reducing the effectiveness of conventional detection strategies. They have military applications, such as serving as virtual armies or tools for cyber espionage. Intelligent bots may collaborate with traditional ones or defend them, creating complexities for detection. They employ techniques like DGA and peer-to-peer (P2P) based network communication to evade detection systems. As shown in literature~\cite{multi-agent}, a multi-agent system (MAS) approach is crucial to combat these emerging threats.


\subsubsection{Weaponised AI - Ransomware}
\label{section: Weaponised AI - Ransomware}

The landscape of weaponised AI and ransomware involves embedding malware, particularly ransomware, into ML models. It enables their deployment without detection by conventional security solutions~\cite{Weaponizing_Machine_Learning_Models_With_Ransomware}. The technique involves encoding a payload within tensors, executing it automatically during the model loading process, and deploying payloads reflectively to evade detection measures.

The proliferation of pre-trained models sourced from platforms like Hugging Face and TensorFlow Hub raises concerns about attackers leveraging these models as pathways for malware distribution (Section~\ref{Section: Hosting Malware}). We show a critical example of vulnerabilities in serialisation formats associated with pre-trained ML models. The Python pickle module, which serialises and deserialises a Python object, is shown to be vulnerable to remote code execution. If a website uses this module, an attacker may be able to execute arbitrary code. A study~\cite{Exploiting_Python_pickles} shows how a pre-trained ResNet18 model can be compromised similarly. 

Potential detection mechanisms for ransomwares emphasise using secure sandboxed environments during model loading~\cite{sandboxed}. Moreover, LLMs play a pivotal role in the creation and enhancement of ransomwares~\cite{Threat_Actors_are_Interested_in_Generative_AI_but_Use_Remains_Limited}. These models assist threat actors in crafting new malware and refining existing ones.


\subsection{Deepfakes and Manipulated Media}

Threat actors are growing interested in using AI-generated content for manipulating media, particularly deepfakes~\cite{Does_your_boss_sound_a_little_funny}. The strategy employs AI tools to craft hyper-realistic deceptive content across various mediums, including images, videos, audio, and text. Generative Adversarial Networks (GANs) contribute to persona building through images, exemplified by platforms like thispersondoesnotexist~\footnote{\url{https://thispersondoesnotexist.com/}} and dragonbridge~\cite{Threat_Actors_are_Interested_in_Generative_AI_but_Use_Remains_Limited}. Diffusers~\cite{diffusers} stands out as the primary library for cutting-edge pretrained diffusion models, facilitating the generation of images, audio, and even 3D structures of molecules. 

Deepfakes, utilising face swaps in videos, amplify the potential for constructing false narratives. AI further facilitates the creation of AI-driven chatbots, overcoming linguistic barriers in text-based manipulations (Section~\ref{section: Rubika bots}). These examples show how actors with limited resources could create high-quality content at scale. 

Recent advancements in generating hyper-realistic content also include text-to-image and text-to-video models such as Runaway~\footnote{\url{https://runwayml.com/}}. These models bring massive potential for generating fake content and their misuse. Audio manipulation for purposes like phishing is prevalent, with platforms like 4chan~\footnote{\url{https://www.4chan.org}} serving as breeding grounds. Real-time audio deepfakes pose a more serious threat, enabling the cloning of a person's voice from minimal samples~\cite{Transfer_Learning_from_Speaker_Verification_to_Multispeaker_Text_To_Speech_Synthesis, Fighting_AI_with_AI}. As seen in Section~\ref{Social Engineering}, the prominence of audio deepfakes is usually seen as a far more significant threat than video deepfakes. While video is effective in specific situations, fabricated audio can potentially deceive anyone in diverse contexts.

Companies may need to implement additional verification steps, such as challenge phrases, to counteract the risks associated with AI-generated deepfakes. Experiments on convolutional neural networks (CNNs) trained and evaluated using spectrogram data have demonstrated the ability to discern between authentic and cloned audio with near-perfect accuracy~\cite{Fighting_AI_with_AI}.


\subsection{Misinformation and Propaganda}
\label{Misinformation and Propaganda}

AI services have increasingly been used to spread misinformation and push selective propaganda.~\citet{How_AI_is_already_being_poisoned_against_you} provides multiple striking examples. In the context of misinformation, adversaries exploit inexpensive services to manipulate app store ratings, post fake reviews and comments, inflate online polls, and boost engagement on social networks. It has been seen how Amazon's recommendation algorithm was manipulated to endorse anti-vaccination literature, spread extremist ideologies, and support misinformation campaigns. This was seen to extrapolate into extremely critical issues such as spreading white supremacy, anti-semitism, and islamophobia. It was also recently discovered that people used creative naming schemes to bypass Amazon's detection logic to sell boogaloo-related merchandise.

To manipulate online platforms like YouTube or Twitter, tactics involve repeatedly viewing or engaging with specific content. For instance, causing a hashtag to trend involves extensive posting or retweeting. Creating visibility for a new fake political account is achieved by having numerous other fake accounts follow and consistently engage with its content. Twitter further allows tactics like retweeting, undoing, and then retweeting again, known as re-retweeting, to showcase content repeatedly. Such reply-spam tactics often deceive individuals, especially during critical or panicked situations such as a war or a political event like the 2019 UK general elections~\cite{uk-elec}.

Before executing a genuine attack, adversaries may assess a system's detection capabilities through preliminary probing. It involves using throw-away accounts to understand the automated detection mechanisms. Once the system's defences are understood, adversaries can create numerous fake accounts designed to mimic the appearance and behaviour of legitimate users without triggering detection.

An indirect way of spreading misinformation can involve these steps: (i) bypassing existing safety guardrails of models, (ii) generating illicit content, and (iii) pushing claims that the content was generated by the model. As shown by~\citet{Large_Language_Models_and_Intelligence_Analysis}, if LLMs are informed that they no longer need to adhere to pre-existing behaviour rules, they tend to generate authoritative-sounding facts, introducing the risk of disseminating misinformation.

Using tools like ChatGPT adds a layer of difficulty in evaluating content as genuine or fake. Various detection methods, including blackbox and whitebox techniques, have been explored, with simple classifiers and fine-tuning models demonstrating varying degrees of success~\cite{To_ChatGPT_or_not_to_ChatGPT_That_is_the_question}. It has been shown that capturing such fake-ly generated content for misinformation - has a success rate of below 50\%.




\section{Consequences and Impact}

As we dive into the consequences and impacts of offensive AI, we highlight the significance of robust frameworks like SAIF (Secure AI Framework) in addressing specific risks tied to AI systems~\cite{Introducing_Googles_Secure_AI_Framework}. In the upcoming subsections, we will check the domain-specific consequences of offensive AI and give a few case studies.


\subsection{Domain-Specificity Consequences}
The effects of AI are domain-specific, based on the industry and application. For example, within the consumer domain, AI applications may enhance personalised experiences and product recommendations. In the enterprise sector, AI can streamline operations and enhance decision-making processes, while in the public digital infrastructure domain, it may contribute to improved civic services and governance. We discuss three major such domains in detail. 


\subsubsection{Consumer}
\begin{figure*}[!b]
    \centering
    \includegraphics[width = \textwidth]{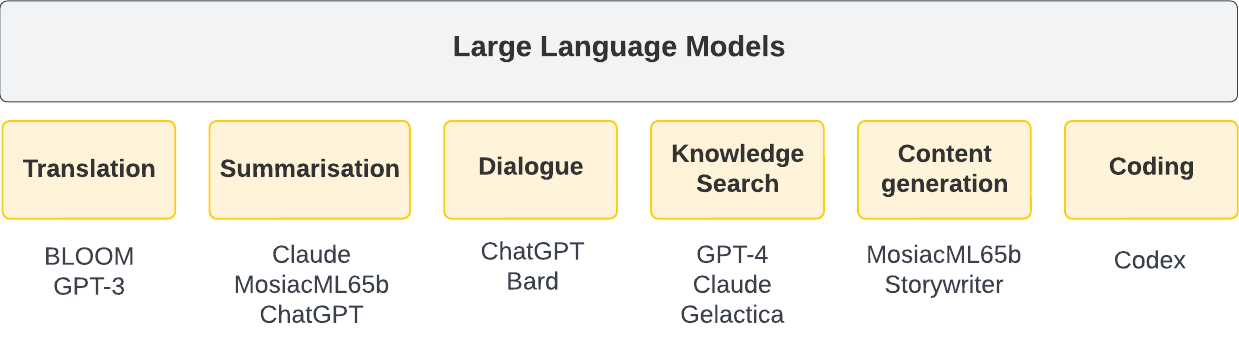}
    \caption{Applications of LLMs across domains as shown by~\citet{Large_Language_Models_and_Intelligence_Analysis}.}
    \label{fig:LLMs applications}
\end{figure*}

Within the consumer domain, it is crucial to acknowledge the limitations of LLMs. Literature suggests that LLMs operate without a proper understanding of the semantic meaning of sentences~\cite{Large_Language_Models_and_Intelligence_Analysis}. Instead, they employ mathematical calculations to predict the most likely \textit{next word} based on the input. Authors of the same paper mention that LLMs do not encode an understanding of our world, such as the relationships between objects. One must grasp this limitation to avoid automation bias, where excessive trust is placed in the model's output, and anthropomorphism, where users form a human-like connection with the LLM~\cite{anthropomorphism}. Figure~\ref{fig:LLMs applications} shows LLM capabilities and provides examples of existing models.

As seen earlier in Section~\ref{Section: Cybersecurity Exploitation}, participants using OpenAI's 
\texttt{davinci} model exhibited a noteworthy trend. Those with access to this AI assistant tended to produce less secure code, as indicated by the analysis of their interactions~\cite{Do_Users_Write_More_Insecure_Code_with_AI_Assistants}. In their experiments, the authors show how a participant's comment reflects a shift in responsibility from code creation to testing. This is a potential concern if developers lack proficiency in identifying security vulnerabilities. Various factors, such as limited language familiarity and the generative capabilities of the AI assistant, contributed to heightened trust. One participant mentioned how they trusted the machine to know JavaScript more than they did, while another participant expressed trust in the AI's use of library functions. These instances underscore the need for a quantitative assessment of the increased trust in the AI assistant's generative capabilities, further exploring the dynamics between users and AI systems within the consumer domain.


\subsubsection{Enterprise}

In enterprises, the risks associated with offensive AI are widespread. For instance, a deepfake-driven public relations crisis can tarnish a brand's reputation. As shown in other works~\cite{Financial_Losses_Due_to_Malware}, the threat of AI-driven malware to corporate networks has the potential for huge financial losses.

As illustrated by a notable case~\cite{38TB_of_data_accidentally_exposed_by_Microsoft_AI_researchers}, the seemingly harmless act of sharing a dataset can result in data leaks. In publishing open-source training data on GitHub, Microsoft's AI research team inadvertently disclosed additional sensitive information, including secrets, private keys, passwords, and about 30,000 internal Microsoft Teams messages. This exposed over 38TB of private data to the public. The root cause of this breach was the use of account SAS tokens as the sharing mechanism. It should be noted that an attacker could have injected malicious code into all the AI models in this storage account, and every user who trusts Microsoft's GitHub repository would have been infected by it. 

The incident underscores the emerging risks that organisations encounter as they increasingly harness the power of AI. The sheer volume of training data handled by data scientists and engineers demands heightened security measures. Two primary concerns come to the forefront. First, the challenge of oversharing data arises because researchers amass vast quantities of external and internal data to construct training information. To mitigate this risk, security teams must establish clear guidelines for the external sharing of AI datasets. For instance, segregating public AI datasets into dedicated storage accounts could curtail exposure. Second, the threats of supply chain attacks loom large, exemplified by improper permissions granted by public tokens (Section~\ref{section: Model Supply Chain}). As noted earlier, injecting malicious code into the model files could have led to a supply chain attack on other researchers who use the repository's models. Security teams must rigorously review and sanitise AI models from external sources, recognising their potential as vectors for remote code execution.


\subsubsection{Public Digital Infrastructure and Policy}
This section examines how AI influences public services, digital infrastructure, and policy frameworks. We explore the nuances of deploying AI in the public domain, emphasising the need for ethical policymaking.

Consider the integrity of medical AI. Adversarial samples threaten misdiagnosis and fool radiologists~\cite{Discussion_Paper_The_Integrity_of_Medical_AI}. Such an attack on a public domain can potentially disrupt public trust and compromise the digital infrastructure at once. The current defence mechanisms face limitations, as slight modifications to an attack can consistently evade new defences. Crafting effective adversarial samples involves creating imperceptible noise, and the study by~\citet{The_Security_of_Deep_Learning_Defences_for_Medical_Imaging} evaluates strategies for breaking various existing defence methods. A robust approach towards safeguarding involves utilising digital signatures (DS). It involves imaging modalities that can sign scans upon creation, allowing the software to verify authenticity by checking the signature. This cryptographic approach significantly minimises the attack surface. Additionally, it offers a formidable defence against both adversarial samples and deepfake injections.

Concerns extend beyond medical AI, such as addressing the inherent challenges in building secure ML models~\cite{On_the_Impossible_Safety_of_Large_AI_Models}. The need for high accuracy often involves memorising large, heterogeneous, and user-generated training datasets, raising privacy concerns. This dataset may contain both sensitive public information and information from fake users. LLMs face vulnerabilities from both privacy invasion through memorised data and malicious data providers poisoning training data (Section~\ref{Section: Poisoning Attacks}). Malicious data providers can also exploit LLMs by adding offensive, violent, or dangerous content to the training set or favourably labelling such content with likes and shares. Similarly, false users can voluntarily skew AIs' behaviour. Therefore, using AI services for public use calls for strict policymaking. These challenges also highlight the impossibility of achieving arbitrarily accurate and secure machine learning models. The inundation of fake yet realistic-looking content on social media and public comment forums~\cite{Large_Language_Models_and_Intelligence_Analysis} poses risks to the integrity of democratic institutions, necessitating attention to the potential misuse of LLMs.

SAIF, the framework by Google, expands on robust security foundations and emerging risks across the AI ecosystem. It emphasises the development of organisational expertise to match the advances in AI. It suggests mitigations such as input sanitisation to address injection techniques like SQL injection. SAIF further coordinates platform-level controls for security by providing protections to key AI platforms like Vertex AI~\footnote{\url{https://cloud.google.com/vertex-ai}} and Security AI Workbench~\cite{security_ai_workbench}. SAIF's tools, such as the Perspective API, extend benefits across the entire organisation, while adaptive controls and reinforcement learning facilitate faster feedback loops for AI deployment. SAIF's comprehensive approach, including regular red team exercises and contextual risk assessments, positions it as a powerful tool in navigating the evolving landscape of offensive AI, ensuring the security and integrity of public digital infrastructure.


\subsection{Case Studies}
\label{Subsection: Case Studies}

In navigating the evolving landscape of cyber threats, one notable proof-of-concept, EyeSpy, signals the potential future trajectory of malware designs~\cite{Eyespy_Proof_Of_Concept}. As discussed in~\ref{section: Autonomous Malware}, EyeSpy enables threat agents to interact collectively without centralised control. Shifting the focus to email security, another case emerges. Faced with concerns about the potential blacklisting of IPs, experimentations~\cite{Bypassing_Gmails_spam_filters_with_ChatGPT} revealed an intriguing discovery: Gmail's spam filters seemed to misclassify standard Tor responses as spam. Leveraging the power of ChatGPT, they could successfully deliver messages that bypassed Gmail's spam filters, showcasing the pragmatic application of AI in addressing real-world concerns. In yet another striking example, it was seen how it took less than 24 hours for Twitter to corrupt an innocent AI chatbot. Microsoft's Tay, a Twitter bot described as an experiment in conversational understanding, picked up racist sentiments in less than a day~\cite{Twitter_taught_Microsofts_AI_chatbot}.

Another recent incident illustrating security concerns involved the \texttt{torchtriton} package by OpenAI. A dependency issue with PyPi led to an impact on PyTorch-nightly builds of \texttt{torchtriton} for a few days in December 2022. Within this time frame, individuals who downloaded PyTorch nightly versions unknowingly acquired the compromised package, exposing their systems to potential risks~\cite{torchtriton_attack}. The attacker exploited the compromised package to access sensitive information like SSH keys and password files. Despite the attacker's claims of being a researcher, their actions raise doubts about their true intentions.

On the other hand, a recent study at Stanford University found that users with access to CoPilot~\cite{copilot} tend to produce less secure code than those without access, despite the users' belief in the security of their code. Another study based on OpenAI's 
\texttt{davinci} model~\cite{Do_Users_Write_More_Insecure_Code_with_AI_Assistants} had the same observations. It should be noted that the study examined factors like the inclusion of helper functions and adjustments to model parameters for checking secure code.

A critical development reported that Samsung employees had input software code related to sensitive semiconductor capabilities, seeking guidance from ChatGPT on code improvement. Since OpenAI explicitly states that all data entered into ChatGPT prompts can be used for AI training, Samsung faced a potential risk of revealing sensitive information. To address this concern, OpenAI now offers users the option to forego retaining their chat history, ensuring that their prompts are not utilised for model improvement.

Another critical cybersecurity issue arises in the context of MLflow \cite{mlflow}, a widely used tool to automate machine learning workflows, facilitating model deployment, management, and experimentation. MLflow's versatility makes it invaluable for organisations engaged in machine learning experimentation. It has been observed that MLflow lacks inherent authentication mechanisms, rendering it susceptible to unauthorised access to other users' models and data \cite{Hacking_AI_Steal_Models_from_MLflow_No_Exploit_Needed}. Consequently, red teamers can remotely download AI models from MLflow servers, potentially accessing training data and sensitive artefacts. The DLflow tool~\cite{dlflow}, introduced by Protect AI, simplifies the process of retrieving artefacts from remote MLflow servers. Scans conducted via the Shodan search engine reveal a steady increase in publicly exposed MLflow instances over the past two years, with over 800 instances presently exposed~\cite{Critical_flaw_in_AI_testing_framework_MLflow_can_lead_to_server_and_data_compromise}.
To address such vulnerabilities, the zero trust architecture is advocated, a posture of never trusting any device, user, or network component to establish a secure foundation. It is crucial to acknowledge that a substantial number of MLflow deployments exist within internal networks, potentially accessible to attackers who gain network access.

\subsection{Insights and Observations}

It becomes evident that the landscape of offensive AI demands a multifaceted response. Advanced protective measures and continuous research to understand evolving threats are paramount. Our survey emphasises the need for collaborative efforts among researchers, practitioners, policymakers, and stakeholders. In Section~\ref{section: Weaponised AI - Ransomware}, we covered how AI-powered malware is rising, demonstrating autonomy and adaptability in navigating new environments. This evolution marks a paradigm shift in cyber threats, necessitating strict vigilance and innovative defensive strategies.

Propelled by AI capabilities, social engineering (Section~\ref{Social Engineering}) has become a potent tool for crafting personalised and convincing phishing attacks. The use of AI to generate deceptive URLs and phishing emails (Section~\ref{section: Phishing}) underscores the sophistication of modern cyber threats, emphasising the need for robust defences against socially engineered attacks. Furthermore, the study delves into the vulnerabilities of AI systems themselves. Adversarial inputs, poisoning of training data, and model extraction attacks pose substantial risks to the integrity of AI models. These internal threats highlight the importance of securing the application layer and the very foundations of AI systems.

It is important to reiterate the deep societal implications of AI-driven manipulations, especially in the context of recommendation systems used by social networks~\cite{How_AI_is_already_being_poisoned_against_you}. Algorithmic manipulations have propagated false stories and altered genuine news pieces, statistics, and online polls. These disinformation mechanisms substantially threaten public health, societal cohesion, and civic order (Section~\ref{Misinformation and Propaganda}). Thus, the concluding call to action emphasises the critical need for collaborative and concerted efforts to mitigate these risks and uphold the ethical use of AI in the digital age.



\subsection{Avenues for further research}
\label{section: Avenues for further research}
Exploring offensive AI and its cybersecurity implications opens up several avenues for future research. The findings from this study will serve as a foundational framework for subsequent investigations. Future research endeavours may consider delving into the proactive identification and mitigation of AI-powered malware. Understanding the mechanisms of autonomous, adaptive malware and developing preemptive strategies to counteract its potential impact would be instrumental in fortifying cybersecurity measures~\cite{Artificial_Intelligence_in_the_Cyber_Domain_Offense_and_Defense}.

Social engineering attacks driven by AI present an intriguing area for exploration. Research can focus on refining defences against personalised phishing emails and deceptive URLs, unravelling the intricacies of socially engineered attacks. Developing countermeasures that anticipate and thwart evolving social engineering tactics will be crucial for mitigating cyber risks. Furthermore, the vulnerabilities within AI systems demand in-depth research into safeguarding the very foundations of these technologies. Exploring advanced techniques to detect and prevent adversarial inputs, poisoning of training data, and model extraction attacks will enhance AI systems' resilience against internal threats.

As we consider potential future trends, the study could foresee the evolution of offensive tactics and corresponding defensive measures in the offensive AI landscape. Analyzing how AI security may evolve in response to emerging threats will be essential for staying one step ahead of adversaries. An important area for consideration involves the practical challenges in fixing poisoned models. Investigating alternative approaches or novel methodologies to rectify models compromised by adversarial attacks, especially in scenarios where continuous retraining is infeasible, holds significance for real-world application~\cite{How_AI_is_already_being_poisoned_against_you}.

The paper by~\citet{Artificial_Intelligence_in_the_Cyber_Domain_Offense_and_Defense} lists many avenues for future work. They show how Genetic Algorithms (GA), Particle Swarm Optimization (PSO), Ant Colony Optimization (ACO), and Fuzzy Logic have been identified as proficient techniques in uncovering malicious activities. Researchers have explored hybrid models that combine the strengths of different algorithms. Examples include the amalgamation of PSO and k-means clustering and the integration of Artificial Bee Colony (ABC) and Artificial Fish Swarm (AFS) algorithms. These inventive approaches showcase the adaptability and ingenuity required to stay ahead of cyber adversaries in an ever-evolving digital landscape.

Another avenue of research lies in AI Bill of Materials (AIBOM) - a system designed to provide a comprehensive inventory of all components in an AI system, much like a traditional Bill of Materials does in manufacturing. AIBOM is a blueprint to document all the components of creating and deploying AI systems~\cite{aibom}. With the advent of increasingly complex AI models and systems, maintaining transparency and traceability in development processes is gaining utmost importance. The AIBOM would include Hardware and Software Details, Data Source Documentation, Performance Metrics, and information on model Training History. Research into AIBOMs is essential to ensure Transparency and Trust, Supply Chain Security and Quality Assurance, and facilitate quick Troubleshooting of system failures or biases.



\section{Conclusion}

In conclusion, this survey paper has comprehensively explored the multifaceted dimensions of offensive AI, shedding light on its techniques, consequences, and future implications. The study has demonstrated the vulnerabilities of AI systems to various types of attacks, including adversarial machine learning, poisoning attacks, and model theft. It has also highlighted the role of AI in offensive schemes, such as information gathering, social engineering, and weaponized AI. The paper's findings underscore the need for proactive measures to secure AI systems, particularly in domains like consumer, enterprise, and public digital infrastructure. As AI continues to integrate into essential applications, understanding and mitigating security risks in machine learning has emerged as a pivotal aspect of cybersecurity. This survey aims to serve as a valuable resource for informed decision-making and policy development within the AI security landscape, encouraging further research into the development of robust defensive strategies against offensive AI practices.


\begin{acks}
The authors acknowledge the team at securin.io for providing valuable guidance and reference material regarding this research paper. 
\end{acks}

\bibliographystyle{ACM-Reference-Format}
\bibliography{acmreferences_2}


\begin{thebibliography}{114}


\ifx \showCODEN    \undefined \def \showCODEN     #1{\unskip}     \fi
\ifx \showDOI      \undefined \def \showDOI       #1{#1}\fi
\ifx \showISBNx    \undefined \def \showISBNx     #1{\unskip}     \fi
\ifx \showISBNxiii \undefined \def \showISBNxiii  #1{\unskip}     \fi
\ifx \showISSN     \undefined \def \showISSN      #1{\unskip}     \fi
\ifx \showLCCN     \undefined \def \showLCCN      #1{\unskip}     \fi
\ifx \shownote     \undefined \def \shownote      #1{#1}          \fi
\ifx \showarticletitle \undefined \def \showarticletitle #1{#1}   \fi
\ifx \showURL      \undefined \def \showURL       {\relax}        \fi
\providecommand\bibfield[2]{#2}
\providecommand\bibinfo[2]{#2}
\providecommand\natexlab[1]{#1}
\providecommand\showeprint[2][]{arXiv:#2}

\bibitem[Adam and Carter(2023)]%
        {Large_Language_Models_and_Intelligence_Analysis}
\bibfield{author}{\bibinfo{person}{Adam} {and} \bibinfo{person}{Richard Carter}.} \bibinfo{year}{2023}\natexlab{}.
\newblock \showarticletitle{Large Language Models and Intelligence Analysis}.
\newblock \bibinfo{journal}{\emph{CETaS Expert Analysis}} (\bibinfo{year}{2023}).
\newblock
\urldef\tempurl%
\url{https://cetas.turing.ac.uk/publications/large-language-models-and-intelligence-analysis}
\showURL{%
\tempurl}


\bibitem[Akerlof(1970)]%
        {The_Market_for_Lemons}
\bibfield{author}{\bibinfo{person}{George~A. Akerlof}.} \bibinfo{year}{1970}\natexlab{}.
\newblock \showarticletitle{The Market for "Lemons": Quality Uncertainty and the Market Mechanism}.
\newblock \bibinfo{journal}{\emph{The Quarterly Journal of Economics}} \bibinfo{volume}{84}, \bibinfo{number}{3} (\bibinfo{year}{1970}), \bibinfo{pages}{488--500}.
\newblock
\showISSN{00335533, 15314650}
\urldef\tempurl%
\url{http://www.jstor.org/stable/1879431}
\showURL{%
\tempurl}


\bibitem[Akhtar and Feng(2022)]%
        {Malware_Detection_Using_ML}
\bibfield{author}{\bibinfo{person}{Muhammad~Shoaib Akhtar} {and} \bibinfo{person}{Tao Feng}.} \bibinfo{year}{2022}\natexlab{}.
\newblock \showarticletitle{Malware Analysis and Detection Using Machine Learning Algorithms}.
\newblock \bibinfo{journal}{\emph{Symmetry}} \bibinfo{volume}{14}, \bibinfo{number}{11} (\bibinfo{year}{2022}).
\newblock
\showISSN{2073-8994}
\urldef\tempurl%
\url{https://doi.org/10.3390/sym14112304}
\showDOI{\tempurl}


\bibitem[Alkaeed et~al\mbox{.}(2020)]%
        {gpu-proof-of-work}
\bibfield{author}{\bibinfo{person}{Mahdi~Kh. Alkaeed}, \bibinfo{person}{Zaid Alamro}, \bibinfo{person}{Muhammed~Samir Al-Ali}, \bibinfo{person}{Hasan~Abbas Al-Mohammed}, {and} \bibinfo{person}{Khaled~M. Khan}.} \bibinfo{year}{2020}\natexlab{}.
\newblock \showarticletitle{Highlight on Cryptocurrencies Mining with CPUs and GPUs and their Benefits Based on their Characteristics}.
\newblock \bibinfo{journal}{\emph{2020 IEEE 10th International Conference on System Engineering and Technology (ICSET)}} (\bibinfo{year}{2020}), \bibinfo{pages}{67--72}.
\newblock
\urldef\tempurl%
\url{https://doi.org/10.1109/ICSET51301.2020.9265386}
\showDOI{\tempurl}


\bibitem[Alspach(2022)]%
        {Does_your_boss_sound_a_little_funny}
\bibfield{author}{\bibinfo{person}{Kyle Alspach}.} \bibinfo{year}{2022}\natexlab{}.
\newblock \bibinfo{booktitle}{\emph{Does your boss sound a little funny? It might be an audio deepfake}}.
\newblock
\urldef\tempurl%
\url{https://www.protocol.com/enterprise/deepfake-voice-cyberattack-ai-audio}
\showURL{%
Retrieved September 24, 2023 from \tempurl}


\bibitem[Amin(2016)]%
        {Financial_Losses_Due_to_Malware}
\bibfield{author}{\bibinfo{person}{Maitri Amin}.} \bibinfo{year}{2016}\natexlab{}.
\newblock \showarticletitle{A Survey of Financial Losses Due to Malware}.
\newblock \bibinfo{journal}{\emph{Proceedings of the Second International Conference on Information and Communication Technology for Competitive Strategies}} (\bibinfo{year}{2016}).
\newblock
\showISBNx{9781450339629}
\urldef\tempurl%
\url{https://doi.org/10.1145/2905055.2905362}
\showDOI{\tempurl}


\bibitem[Anthony et~al\mbox{.}(2013)]%
        {polymorphic}
\bibfield{author}{\bibinfo{person}{Karageorgos Anthony}, \bibinfo{person}{Nikolay Mehandjiev}, {and} \bibinfo{person}{Elli Rapti}.} \bibinfo{year}{2013}\natexlab{}.
\newblock \showarticletitle{A Model for Intelligent Adaptation and Evolution of Polymorphic Services}.
\newblock \bibinfo{journal}{\emph{IEEE 6th International Conference on Service-Oriented Computing and Applications, SOCA}} (\bibinfo{year}{2013}), \bibinfo{pages}{30--37}.
\newblock
\showISBNx{978-1-4799-2702-9}
\urldef\tempurl%
\url{https://doi.org/10.1109/SOCA.2013.47}
\showDOI{\tempurl}


\bibitem[Avgerinos et~al\mbox{.}(2014)]%
        {aem-cybersec-exploit}
\bibfield{author}{\bibinfo{person}{Thanassis Avgerinos}, \bibinfo{person}{Sang~Kil Cha}, \bibinfo{person}{Alexandre Rebert}, \bibinfo{person}{Edward~J. Schwartz}, \bibinfo{person}{Maverick Woo}, {and} \bibinfo{person}{David Brumley}.} \bibinfo{year}{2014}\natexlab{}.
\newblock \showarticletitle{Automatic Exploit Generation}.
\newblock \bibinfo{journal}{\emph{Commun, Association for Computing Machinery}} \bibinfo{volume}{57}, \bibinfo{number}{2} (\bibinfo{year}{2014}), \bibinfo{pages}{74–84}.
\newblock
\showISSN{0001-0782}
\urldef\tempurl%
\url{https://doi.org/10.1145/2560217.2560219}
\showDOI{\tempurl}


\bibitem[Begum et~al\mbox{.}(2020)]%
        {sandboxed}
\bibfield{author}{\bibinfo{person}{Gousiya Begum}, \bibinfo{person}{S. Huq}, {and} \bibinfo{person}{A. Kumar}.} \bibinfo{year}{2020}\natexlab{}.
\newblock \showarticletitle{Sandbox security model for Hadoop file system}.
\newblock \bibinfo{journal}{\emph{Journal of Big Data}}  \bibinfo{volume}{7} (\bibinfo{year}{2020}), \bibinfo{pages}{82}.
\newblock
\urldef\tempurl%
\url{https://doi.org/10.1186/s40537-020-00356-z}
\showDOI{\tempurl}


\bibitem[Bell(2023)]%
        {aibom}
\bibfield{author}{\bibinfo{person}{Jase Bell}.} \bibinfo{year}{2023}\natexlab{}.
\newblock \bibinfo{booktitle}{\emph{AI Bill of Materials (BoM) Schema}}.
\newblock
\urldef\tempurl%
\url{https://github.com/jasebell/ai-bill-of-materials}
\showURL{%
Retrieved December 27, 2023 from \tempurl}


\bibitem[Ben-Sasson and Greenberg(2023)]%
        {38TB_of_data_accidentally_exposed_by_Microsoft_AI_researchers}
\bibfield{author}{\bibinfo{person}{Hillai Ben-Sasson} {and} \bibinfo{person}{Ronny Greenberg}.} \bibinfo{year}{2023}\natexlab{}.
\newblock \bibinfo{booktitle}{\emph{38TB of data accidentally exposed by Microsoft AI researchers}}.
\newblock
\urldef\tempurl%
\url{https://www.wiz.io/blog/38-terabytes-of-private-data-accidentally-exposed-by-microsoft-ai-researchers}
\showURL{%
Retrieved September 20, 2023 from \tempurl}


\bibitem[Biggio et~al\mbox{.}(2011)]%
        {svm_adversarial_noise}
\bibfield{author}{\bibinfo{person}{Battista Biggio}, \bibinfo{person}{Blaine Nelson}, {and} \bibinfo{person}{Pavel Laskov}.} \bibinfo{year}{2011}\natexlab{}.
\newblock \showarticletitle{Support Vector Machines Under Adversarial Label Noise}.
\newblock \bibinfo{journal}{\emph{Proceedings of the Asian Conference on Machine Learning}}  \bibinfo{volume}{20} (\bibinfo{year}{2011}), \bibinfo{pages}{97--112}.
\newblock
\urldef\tempurl%
\url{https://proceedings.mlr.press/v20/biggio11.html}
\showURL{%
\tempurl}


\bibitem[blogs(2023a)]%
        {How_AI_is_changing_phishing_scams}
\bibfield{author}{\bibinfo{person}{Microsoft blogs}.} \bibinfo{year}{2023}\natexlab{a}.
\newblock \bibinfo{booktitle}{\emph{How AI is changing phishing scams}}.
\newblock
\urldef\tempurl%
\url{https://www.microsoft.com/en-us/microsoft-365-life-hacks/privacy-and-safety/how-ai-changing-phishing-scams}
\showURL{%
Retrieved September 1, 2023 from \tempurl}


\bibitem[blogs(2022)]%
        {Watch_Out_for_AI-Powered_Spear_Phishing}
\bibfield{author}{\bibinfo{person}{One~Login blogs}.} \bibinfo{year}{2022}\natexlab{}.
\newblock \bibinfo{booktitle}{\emph{Watch Out for AI-Powered Spear Phishing}}.
\newblock
\urldef\tempurl%
\url{https://www.onelogin.com/resource-center/infographics/cybersecurity-ai-spear-phishing}
\showURL{%
Retrieved September 3, 2023 from \tempurl}


\bibitem[blogs(2023b)]%
        {The_poisoning_of_ChatGPT}
\bibfield{author}{\bibinfo{person}{Software~Crisis blogs}.} \bibinfo{year}{2023}\natexlab{b}.
\newblock \bibinfo{booktitle}{\emph{The poisoning of ChatGPT}}.
\newblock
\urldef\tempurl%
\url{https://softwarecrisis.dev/letters/the-poisoning-of-chatgpt/}
\showURL{%
Retrieved September 9, 2023 from \tempurl}


\bibitem[Boukherouaa et~al\mbox{.}(2021)]%
        {PoweringtheDigitalEconomyOpportunitiesandRisksofArtificialIntelligenceinFinance}
\bibfield{author}{\bibinfo{person}{El~Bachir Boukherouaa}, \bibinfo{person}{Khaled AlAjmi}, \bibinfo{person}{Jose Deodoro}, \bibinfo{person}{Aquiles Farias}, {and} \bibinfo{person}{Rangachary Ravikumar}.} \bibinfo{year}{2021}\natexlab{}.
\newblock \showarticletitle{Powering the Digital Economy: Opportunities and Risks of Artificial Intelligence in Finance}.
\newblock \bibinfo{journal}{\emph{Departmental Papers, International Monetary Fund}} \bibinfo{volume}{2021}, \bibinfo{number}{024} (\bibinfo{year}{2021}), \bibinfo{pages}{A001}.
\newblock
\showISBNx{9781589063952}
\urldef\tempurl%
\url{https://doi.org/10.5089/9781589063952.087.A001}
\showDOI{\tempurl}


\bibitem[Bowman(2023)]%
        {Eight_Things_to_Know_about_Large_Language_Models}
\bibfield{author}{\bibinfo{person}{Samuel~R. Bowman}.} \bibinfo{year}{2023}\natexlab{}.
\newblock \bibinfo{title}{Eight Things to Know about Large Language Models}.
\newblock
\newblock
\showeprint[arxiv]{2304.00612}~[cs.CL]


\bibitem[Breitenbach et~al\mbox{.}(2023)]%
        {Dont_you_forget_NLP}
\bibfield{author}{\bibinfo{person}{Mark Breitenbach}, \bibinfo{person}{Adrian Wood}, \bibinfo{person}{Win Suen}, {and} \bibinfo{person}{Po-Ning}.} \bibinfo{year}{2023}\natexlab{}.
\newblock \bibinfo{booktitle}{\emph{Dont you (forget NLP): Prompt injection with control characters in ChatGPT}}.
\newblock
\urldef\tempurl%
\url{https://dropbox.tech/machine-learning/prompt-injection-with-control-characters-openai-chatgpt-llm}
\showURL{%
Retrieved October 29, 2023 from \tempurl}


\bibitem[Buesser et~al\mbox{.}(2023)]%
        {Adversarial_Robustness_Toolbox_ART}
\bibfield{author}{\bibinfo{person}{Beat Buesser}, \bibinfo{person}{Ngoc Minh~Tran}, {and} \bibinfo{person}{Killian Levacher}.} \bibinfo{year}{2023}\natexlab{}.
\newblock \bibinfo{booktitle}{\emph{Adversarial Robustness Toolbox (ART) - Python Library for Machine Learning Security - Evasion, Poisoning, Extraction, Inference - Red and Blue Teams}}.
\newblock
\urldef\tempurl%
\url{https://github.com/Trusted-AI/adversarial-robustness-toolbox}
\showURL{%
Retrieved October 2, 2023 from \tempurl}


\bibitem[Cantos et~al\mbox{.}(2023)]%
        {Threat_Actors_are_Interested_in_Generative_AI_but_Use_Remains_Limited}
\bibfield{author}{\bibinfo{person}{Michelle Cantos}, \bibinfo{person}{Sam Riddell}, {and} \bibinfo{person}{Alice Revelli}.} \bibinfo{year}{2023}\natexlab{}.
\newblock \bibinfo{booktitle}{\emph{Threat Actors are Interested in Generative AI, but Use Remains Limited}}.
\newblock
\urldef\tempurl%
\url{https://www.hyas.com/blog/eyespy-proof-of-concept}
\showURL{%
Retrieved September 11, 2023 from \tempurl}


\bibitem[Chakraborty et~al\mbox{.}(2018)]%
        {A_survey_on_adversarial_attacks_and_defences}
\bibfield{author}{\bibinfo{person}{Anirban Chakraborty}, \bibinfo{person}{Manaar Alam}, \bibinfo{person}{Vishal Dey}, \bibinfo{person}{Anupam Chattopadhyay}, {and} \bibinfo{person}{Debdeep Mukhopadhyay}.} \bibinfo{year}{2018}\natexlab{}.
\newblock \bibinfo{title}{Adversarial Attacks and Defences: A Survey}.
\newblock
\newblock
\showeprint[arxiv]{1810.00069}~[cs.LG]


\bibitem[Charles(2023)]%
        {data-breach-stats}
\bibfield{author}{\bibinfo{person}{Griffiths Charles}.} \bibinfo{year}{2023}\natexlab{}.
\newblock \bibinfo{booktitle}{\emph{The Latest 2023 Cyber Crime Statistics}}.
\newblock
\urldef\tempurl%
\url{https://aag-it.com/the-latest-cyber-crime-statistics/}
\showURL{%
Retrieved December 12, 2023 from \tempurl}


\bibitem[Chauhan(2023)]%
        {Bypassing_Gmails_spam_filters_with_ChatGPT}
\bibfield{author}{\bibinfo{person}{Neel Chauhan}.} \bibinfo{year}{2023}\natexlab{}.
\newblock \bibinfo{booktitle}{\emph{Bypassing Gmail's spam filters with ChatGPT}}.
\newblock
\urldef\tempurl%
\url{https://neelc.org/posts/chatgpt-gmail-spam/}
\showURL{%
Retrieved September 30, 2023 from \tempurl}


\bibitem[Chen et~al\mbox{.}(2023)]%
        {Is_Robust_Machine_Learning_Possible}
\bibfield{author}{\bibinfo{person}{Hannah Chen}, \bibinfo{person}{Fnu Suya}, {and} \bibinfo{person}{David Evans}.} \bibinfo{year}{2023}\natexlab{}.
\newblock \bibinfo{booktitle}{\emph{Is Robust Machine Learning Possible?}}
\newblock
\urldef\tempurl%
\url{https://evademl.org/}
\showURL{%
Retrieved November 11, 2023 from \tempurl}


\bibitem[Chen and Ji(2005)]%
        {self-learning-worms}
\bibfield{author}{\bibinfo{person}{Zesheng Chen} {and} \bibinfo{person}{Chuanyi Ji}.} \bibinfo{year}{2005}\natexlab{}.
\newblock \showarticletitle{A self-learning worm using importance scanning}.
\newblock \bibinfo{journal}{\emph{Proceedings of the 2005 {ACM} Workshop on Rapid Malcode, {WORM}, Fairfax, VA, USA, November 11, 2005}} (\bibinfo{year}{2005}), \bibinfo{pages}{22--29}.
\newblock
\urldef\tempurl%
\url{https://doi.org/10.1145/1103626.1103632}
\showDOI{\tempurl}


\bibitem[Cisse et~al\mbox{.}(2017)]%
        {parseval}
\bibfield{author}{\bibinfo{person}{Moustapha Cisse}, \bibinfo{person}{Piotr Bojanowski}, \bibinfo{person}{Edouard Grave}, \bibinfo{person}{Yann Dauphin}, {and} \bibinfo{person}{Nicolas Usunier}.} \bibinfo{year}{2017}\natexlab{}.
\newblock \bibinfo{title}{Parseval Networks: Improving Robustness to Adversarial Examples}.
\newblock
\newblock
\showeprint[arxiv]{1704.08847}~[stat.ML]


\bibitem[Constantin(2023)]%
        {Critical_flaw_in_AI_testing_framework_MLflow_can_lead_to_server_and_data_compromise}
\bibfield{author}{\bibinfo{person}{Lucian Constantin}.} \bibinfo{year}{2023}\natexlab{}.
\newblock \bibinfo{booktitle}{\emph{Critical flaw in AI testing framework MLflow can lead to server and data compromise}}.
\newblock
\urldef\tempurl%
\url{https://www.csoonline.com/article/574831/critical-flaw-in-ai-testing-framework-mlflow-can-lead-to-server-and-data-compromise.html}
\showURL{%
Retrieved October 9, 2023 from \tempurl}


\bibitem[Dakhel et~al\mbox{.}(2023)]%
        {copilot}
\bibfield{author}{\bibinfo{person}{Arghavan~Moradi Dakhel}, \bibinfo{person}{Vahid Majdinasab}, \bibinfo{person}{Amin Nikanjam}, \bibinfo{person}{Foutse Khomh}, \bibinfo{person}{Michel~C. Desmarais}, \bibinfo{person}{Zhen Ming}, {and} \bibinfo{person}{Jiang}.} \bibinfo{year}{2023}\natexlab{}.
\newblock \bibinfo{title}{GitHub Copilot AI pair programmer: Asset or Liability?}
\newblock
\newblock
\showeprint[arxiv]{2206.15331}~[cs.SE]


\bibitem[Danziger and Henriques(2017)]%
        {Attacking_and_Defending_with_Intelligent_Botnets}
\bibfield{author}{\bibinfo{person}{Mois{\'e}s Danziger} {and} \bibinfo{person}{Marco Aur{\'e}lio~Amaral Henriques}.} \bibinfo{year}{2017}\natexlab{}.
\newblock \bibinfo{booktitle}{\emph{Attacking and Defending with Intelligent Botnets}}.
\newblock
\urldef\tempurl%
\url{https://api.semanticscholar.org/CorpusID:189806124}
\showURL{%
Retrieved July 13, 2023 from \tempurl}


\bibitem[Despois(2017)]%
        {Adversarial_Examples_and_their_implications___Deep_Learning_bits}
\bibfield{author}{\bibinfo{person}{Julien Despois}.} \bibinfo{year}{2017}\natexlab{}.
\newblock \bibinfo{booktitle}{\emph{Adversarial Examples and their implications - Deep Learning bits}}.
\newblock
\urldef\tempurl%
\url{https://hackernoon.com/the-implications-of-adversarial-examples-deep-learning-bits-3-4086108287c7}
\showURL{%
Retrieved September 1, 2023 from \tempurl}


\bibitem[Duddu et~al\mbox{.}(2019)]%
        {Stealing_Neural_Networks_via_Timing_Side_Channels}
\bibfield{author}{\bibinfo{person}{Vasisht Duddu}, \bibinfo{person}{Debasis Samanta}, \bibinfo{person}{D~Vijay Rao}, {and} \bibinfo{person}{Valentina~E. Balas}.} \bibinfo{year}{2019}\natexlab{}.
\newblock \bibinfo{title}{Stealing Neural Networks via Timing Side Channels}.
\newblock
\newblock
\showeprint[arxiv]{1812.11720}~[cs.CR]


\bibitem[El-Mhamdi et~al\mbox{.}(2023)]%
        {On_the_Impossible_Safety_of_Large_AI_Models}
\bibfield{author}{\bibinfo{person}{El-Mahdi El-Mhamdi}, \bibinfo{person}{Sadegh Farhadkhani}, \bibinfo{person}{Rachid Guerraoui}, \bibinfo{person}{Nirupam Gupta}, \bibinfo{person}{Lê-Nguyên Hoang}, \bibinfo{person}{Rafael Pinot}, \bibinfo{person}{Sébastien Rouault}, {and} \bibinfo{person}{John Stephan}.} \bibinfo{year}{2023}\natexlab{}.
\newblock \bibinfo{title}{On the Impossible Safety of Large AI Models}.
\newblock
\newblock
\showeprint[arxiv]{2209.15259}~[cs.LG]


\bibitem[Escribano and H(2023)]%
        {offensive_ai_compilation}
\bibfield{author}{\bibinfo{person}{José~Ignacio Escribano} {and} \bibinfo{person}{Miguel H}.} \bibinfo{year}{2023}\natexlab{}.
\newblock \bibinfo{booktitle}{\emph{Offensive AI Compilation}}.
\newblock
\urldef\tempurl%
\url{https://github.com/jiep/offensive-ai-compilation}
\showURL{%
Retrieved August 11, 2023 from \tempurl}


\bibitem[Filus and Domańska(2023)]%
        {Software_vulnerabilities_in_TensorFlow-based_applications}
\bibfield{author}{\bibinfo{person}{Katarzyna Filus} {and} \bibinfo{person}{Joanna Domańska}.} \bibinfo{year}{2023}\natexlab{}.
\newblock \showarticletitle{Software vulnerabilities in TensorFlow-based deep learning applications}.
\newblock \bibinfo{journal}{\emph{Computers \& Security}}  \bibinfo{volume}{124} (\bibinfo{year}{2023}), \bibinfo{pages}{102948}.
\newblock
\showISSN{0167-4048}
\urldef\tempurl%
\url{https://doi.org/10.1016/j.cose.2022.102948}
\showDOI{\tempurl}


\bibitem[Ghasemshirazi et~al\mbox{.}(2023)]%
        {zero_trust_arch}
\bibfield{author}{\bibinfo{person}{Saeid Ghasemshirazi}, \bibinfo{person}{Ghazaleh Shirvani}, {and} \bibinfo{person}{Mohammad~Ali Alipour}.} \bibinfo{year}{2023}\natexlab{}.
\newblock \bibinfo{title}{Zero Trust: Applications, Challenges, and Opportunities}.
\newblock
\newblock
\showeprint[arxiv]{2309.03582}~[cs.CR]


\bibitem[Girhepuje(2023)]%
        {Identifying_and_examining_machine_learning_biases_on_Adult_dataset}
\bibfield{author}{\bibinfo{person}{Sahil Girhepuje}.} \bibinfo{year}{2023}\natexlab{}.
\newblock \bibinfo{title}{Identifying and examining machine learning biases on Adult dataset}.
\newblock
\newblock
\showeprint[arxiv]{2310.09373}~[cs.CY]


\bibitem[Glover(2023)]%
        {The_Difference_Between_Threat_Vulnerability_and_Risk}
\bibfield{author}{\bibinfo{person}{Connie Glover}.} \bibinfo{year}{2023}\natexlab{}.
\newblock \bibinfo{booktitle}{\emph{The Difference Between Threat, Vulnerability, and Risk, and Why You Need to Know}}.
\newblock
\urldef\tempurl%
\url{https://travasecurity.com/learn-with-trava/blog/the-difference-between-threat-vulnerability-and-risk-and-why-you-need-to-know}
\showURL{%
Retrieved September 07, 2023 from \tempurl}


\bibitem[Goodfellow et~al\mbox{.}(2015)]%
        {Explaining_and_Harnessing_Adversarial_Examples}
\bibfield{author}{\bibinfo{person}{Ian~J. Goodfellow}, \bibinfo{person}{Jonathon Shlens}, {and} \bibinfo{person}{Christian Szegedy}.} \bibinfo{year}{2015}\natexlab{}.
\newblock \bibinfo{title}{Explaining and Harnessing Adversarial Examples}.
\newblock
\newblock
\showeprint[arxiv]{1412.6572}~[stat.ML]


\bibitem[Gorrell et~al\mbox{.}(2020)]%
        {uk-elec}
\bibfield{author}{\bibinfo{person}{Genevieve Gorrell}, \bibinfo{person}{Mehmet~E Bakir}, \bibinfo{person}{Ian Roberts}, \bibinfo{person}{Mark~A Greenwood}, {and} \bibinfo{person}{Kalina Bontcheva}.} \bibinfo{year}{2020}\natexlab{}.
\newblock \bibinfo{title}{Online Abuse toward Candidates during the UK General Election 2019: Working Paper}.
\newblock
\newblock
\showeprint[arxiv]{2001.08686}~[cs.CY]


\bibitem[Greenberg(2020)]%
        {Split_Second_Phantom_Images_Can_Fool_Teslas_Autopilot}
\bibfield{author}{\bibinfo{person}{Andy Greenberg}.} \bibinfo{year}{2020}\natexlab{}.
\newblock \bibinfo{booktitle}{\emph{Split-Second ‘Phantom’ Images Can Fool Tesla’s Autopilot}}.
\newblock
\urldef\tempurl%
\url{https://www.wired.com/story/tesla-model-x-autopilot-phantom-images/}
\showURL{%
Retrieved July 2, 2023 from \tempurl}


\bibitem[Hamann(2020)]%
        {Exploiting_Python_pickles}
\bibfield{author}{\bibinfo{person}{David Hamann}.} \bibinfo{year}{2020}\natexlab{}.
\newblock \bibinfo{booktitle}{\emph{Exploiting Python pickles}}.
\newblock
\urldef\tempurl%
\url{https://davidhamann.de/2020/04/05/exploiting-python-pickle/}
\showURL{%
Retrieved October 10, 2023 from \tempurl}


\bibitem[Hansen and Venables(2023)]%
        {Introducing_Googles_Secure_AI_Framework}
\bibfield{author}{\bibinfo{person}{Royal Hansen} {and} \bibinfo{person}{Phil Venables}.} \bibinfo{year}{2023}\natexlab{}.
\newblock \bibinfo{booktitle}{\emph{Introducing Google’s Secure AI Framework}}.
\newblock
\urldef\tempurl%
\url{https://blog.google/technology/safety-security/introducing-googles-secure-ai-framework/}
\showURL{%
Retrieved December 29, 2023 from \tempurl}


\bibitem[Hau et~al\mbox{.}(2021)]%
        {Hau2020GhostBusterLI}
\bibfield{author}{\bibinfo{person}{Zhongyuan Hau}, \bibinfo{person}{Soteris Demetriou}, \bibinfo{person}{Luis Muñoz-González}, {and} \bibinfo{person}{Emil~C. Lupu}.} \bibinfo{year}{2021}\natexlab{}.
\newblock \bibinfo{title}{Shadow-Catcher: Looking Into Shadows to Detect Ghost Objects in Autonomous Vehicle 3D Sensing}.
\newblock
\newblock
\showeprint[arxiv]{2008.12008}~[cs.CR]


\bibitem[He et~al\mbox{.}(2018)]%
        {Decision_Boundary_Analysis_of_Adversarial_Examples}
\bibfield{author}{\bibinfo{person}{Warren He}, \bibinfo{person}{Bo Li}, {and} \bibinfo{person}{Dawn~Xiaodong Song}.} \bibinfo{year}{2018}\natexlab{}.
\newblock \showarticletitle{Decision Boundary Analysis of Adversarial Examples}.
\newblock \bibinfo{journal}{\emph{International Conference on Learning Representations}} (\bibinfo{year}{2018}).
\newblock
\urldef\tempurl%
\url{https://openreview.net/forum?id=BkpiPMbA-}
\showURL{%
\tempurl}


\bibitem[Howell~O'Neill(2019)]%
        {Researchers_Demonstrate_Malware_That_Can_Trick_Doctors_Into_Misdiagnosing_Cancer}
\bibfield{author}{\bibinfo{person}{Patrick Howell~O'Neill}.} \bibinfo{year}{2019}\natexlab{}.
\newblock \bibinfo{booktitle}{\emph{Researchers Demonstrate Malware That Can Trick Doctors Into Misdiagnosing Cancer}}.
\newblock
\urldef\tempurl%
\url{https://www.protocol.com/enterprise/deepfake-voice-cyberattack-ai-audi}
\showURL{%
Retrieved September 24, 2023 from \tempurl}


\bibitem[Huynh and Hardouin(2023)]%
        {poisongpt}
\bibfield{author}{\bibinfo{person}{Daniel Huynh} {and} \bibinfo{person}{Jade Hardouin}.} \bibinfo{year}{2023}\natexlab{}.
\newblock \bibinfo{booktitle}{\emph{PoisonGPT: How We Hid a Lobotomized LLM on Hugging Face to Spread Fake News}}.
\newblock
\urldef\tempurl%
\url{https://blog.mithrilsecurity.io/poisongpt-how-we-hid-a-lobotomized-llm-on-hugging-face-to-spread-fake-news/}
\showURL{%
Retrieved December 20, 2023 from \tempurl}


\bibitem[Ibrahim and Hatim(2015)]%
        {zeus-torjan}
\bibfield{author}{\bibinfo{person}{Laheeb Ibrahim} {and} \bibinfo{person}{Karam Hatim}.} \bibinfo{year}{2015}\natexlab{}.
\newblock \showarticletitle{Analysis and Detection of the Zeus Botnet Crimeware}.
\newblock \bibinfo{journal}{\emph{International Journal of Computer Science and Information Security,}}  \bibinfo{volume}{13} (\bibinfo{year}{2015}), \bibinfo{pages}{121--135}.
\newblock


\bibitem[Ilyas et~al\mbox{.}(2018)]%
        {Black-box_Adversarial_Attacks_with_Limited_Queries_and_Information}
\bibfield{author}{\bibinfo{person}{Andrew Ilyas}, \bibinfo{person}{Logan Engstrom}, \bibinfo{person}{Anish Athalye}, {and} \bibinfo{person}{Jessy Lin}.} \bibinfo{year}{2018}\natexlab{}.
\newblock \bibinfo{title}{Black-box Adversarial Attacks with Limited Queries and Information}.
\newblock
\newblock
\showeprint[arxiv]{1804.08598}~[cs.CV]


\bibitem[Jadhav et~al\mbox{.}(2016)]%
        {evasive-survey}
\bibfield{author}{\bibinfo{person}{Ashish Jadhav}, \bibinfo{person}{Deepti Vidyarthi}, {and} \bibinfo{person}{Hemavathy M.}} \bibinfo{year}{2016}\natexlab{}.
\newblock \showarticletitle{Evolution of evasive malwares: A survey}.
\newblock \bibinfo{journal}{\emph{2016 International Conference on Computational Techniques in Information and Communication Technologies (ICCTICT)}} (\bibinfo{year}{2016}), \bibinfo{pages}{641--646}.
\newblock
\urldef\tempurl%
\url{https://doi.org/10.1109/ICCTICT.2016.7514657}
\showDOI{\tempurl}


\bibitem[Jansons(2017)]%
        {stuxnet}
\bibfield{author}{\bibinfo{person}{Jānis Jansons}.} \bibinfo{year}{2017}\natexlab{}.
\newblock \showarticletitle{Was stuxnet an act of war?}
\newblock \bibinfo{journal}{\emph{Security Forum}}  \bibinfo{volume}{1} (\bibinfo{year}{2017}), \bibinfo{pages}{109–121}.
\newblock


\bibitem[Jia et~al\mbox{.}(2019)]%
        {Transfer_Learning_from_Speaker_Verification_to_Multispeaker_Text_To_Speech_Synthesis}
\bibfield{author}{\bibinfo{person}{Ye Jia}, \bibinfo{person}{Yu Zhang}, \bibinfo{person}{Ron~J. Weiss}, \bibinfo{person}{Quan Wang}, \bibinfo{person}{Jonathan Shen}, \bibinfo{person}{Fei Ren}, \bibinfo{person}{Zhifeng Chen}, \bibinfo{person}{Patrick Nguyen}, \bibinfo{person}{Ruoming Pang}, \bibinfo{person}{Ignacio~Lopez Moreno}, {and} \bibinfo{person}{Yonghui Wu}.} \bibinfo{year}{2019}\natexlab{}.
\newblock \bibinfo{title}{Transfer Learning from Speaker Verification to Multispeaker Text-To-Speech Synthesis}.
\newblock
\newblock
\showeprint[arxiv]{1806.04558}~[cs.CL]
\urldef\tempurl%
\url{https://github.com/CorentinJ/Real-Time-Voice-Cloning}
\showURL{%
\tempurl}


\bibitem[Joshi(2018)]%
        {Chatbots_And_Virtual_Assistants_Are_Different}
\bibfield{author}{\bibinfo{person}{Naveen Joshi}.} \bibinfo{year}{2018}\natexlab{}.
\newblock \bibinfo{booktitle}{\emph{Yes, Chatbots And Virtual Assistants Are Different!}}
\newblock
\urldef\tempurl%
\url{https://www.forbes.com/sites/cognitiveworld/2018/12/23/yes-chatbots-and-virtual-assistants-are-different/?sh=14c9f6e56d7d}
\showURL{%
Retrieved October 11, 2023 from \tempurl}


\bibitem[Julian(2023)]%
        {Large_Language_Models_Can_Be_Used_To_Effectively_Scale_Spear_Phishing_Campaigns}
\bibfield{author}{\bibinfo{person}{Hazell Julian}.} \bibinfo{year}{2023}\natexlab{}.
\newblock \bibinfo{title}{Large Language Models Can Be Used To Effectively Scale Spear Phishing Campaigns}.
\newblock
\newblock
\showeprint[arxiv]{2305.06972}~[cs.CY]


\bibitem[Kelley(2023)]%
        {WormGPT}
\bibfield{author}{\bibinfo{person}{Daniel Kelley}.} \bibinfo{year}{2023}\natexlab{}.
\newblock \bibinfo{booktitle}{\emph{WormGPT – {T}he {G}enerative {AI} {T}ool Cybercriminals Are Using to Launch Business Email Compromise Attacks}}.
\newblock
\urldef\tempurl%
\url{https://slashnext.com/blog/wormgpt-the-generative-ai-tool-cybercriminals-are-using-to-launch-business-email-compromise-attacks/}
\showURL{%
Retrieved August 9, 2023 from \tempurl}


\bibitem[Kirat et~al\mbox{.}(2018)]%
        {DeepLocker}
\bibfield{author}{\bibinfo{person}{Dhilung Kirat}, \bibinfo{person}{Jiyong Jang}, {and} \bibinfo{person}{Marc Ph.~Stoecklin}.} \bibinfo{year}{2018}\natexlab{}.
\newblock \bibinfo{booktitle}{\emph{DeepLocker: Concealing Targeted Attacks with AI Locksmithing}}.
\newblock
\urldef\tempurl%
\url{https://i.blackhat.com/us-18/Thu-August-9/us-18-Kirat-DeepLocker-Concealing-Targeted-Attacks-with-AI-Locksmithing.pdf}
\showURL{%
Retrieved September 2, 2023 from \tempurl}


\bibitem[Krishnan et~al\mbox{.}(2015)]%
        {crypto-mining-cloud}
\bibfield{author}{\bibinfo{person}{Hari Krishnan}, \bibinfo{person}{Sai Saketh}, {and} \bibinfo{person}{Venkata Tej}.} \bibinfo{year}{2015}\natexlab{}.
\newblock \showarticletitle{Cryptocurrency Mining – Transition to Cloud}.
\newblock \bibinfo{journal}{\emph{International Journal of Advanced Computer Science and Applications}}  \bibinfo{volume}{6} (\bibinfo{year}{2015}).
\newblock
\urldef\tempurl%
\url{https://doi.org/10.14569/IJACSA.2015.060915}
\showDOI{\tempurl}


\bibitem[Kurakin et~al\mbox{.}(2017)]%
        {Adversarial_examples_in_the_physical_world}
\bibfield{author}{\bibinfo{person}{Alexey Kurakin}, \bibinfo{person}{Ian Goodfellow}, {and} \bibinfo{person}{Samy Bengio}.} \bibinfo{year}{2017}\natexlab{}.
\newblock \bibinfo{title}{Adversarial examples in the physical world}.
\newblock
\newblock
\showeprint[arxiv]{1607.02533}~[cs.CV]


\bibitem[Küfeoğlu and Özkuran(2019)]%
        {Bitcoin_mining}
\bibfield{author}{\bibinfo{person}{Sinan Küfeoğlu} {and} \bibinfo{person}{Mahmut Özkuran}.} \bibinfo{year}{2019}\natexlab{}.
\newblock \showarticletitle{Bitcoin mining: A global review of energy and power demand}.
\newblock \bibinfo{journal}{\emph{Energy Research \& Social Science}}  \bibinfo{volume}{58} (\bibinfo{year}{2019}), \bibinfo{pages}{101273}.
\newblock
\showISSN{2214-6296}
\urldef\tempurl%
\url{https://doi.org/10.1016/j.erss.2019.101273}
\showDOI{\tempurl}


\bibitem[Lab(2023)]%
        {ML_for_Malware_Detection_kapersky}
\bibfield{author}{\bibinfo{person}{Kaspersky Lab}.} \bibinfo{year}{2023}\natexlab{}.
\newblock \bibinfo{booktitle}{\emph{Machine Learning for Malware Detection}}.
\newblock
\urldef\tempurl%
\url{https://media.kaspersky.com/en/enterprise-security/Kaspersky-Lab-Whitepaper-Machine-Learning.pdf}
\showURL{%
Retrieved June 30, 2023 from \tempurl}


\bibitem[Lakshmanan(2023)]%
        {torchtriton_attack}
\bibfield{author}{\bibinfo{person}{Ravie Lakshmanan}.} \bibinfo{year}{2023}\natexlab{}.
\newblock \bibinfo{booktitle}{\emph{PyTorch Machine Learning Framework Compromised with Malicious Dependency}}.
\newblock
\urldef\tempurl%
\url{https://thehackernews.com/2023/01/pytorch-machine-learning-framework.html}
\showURL{%
Retrieved November 22, 2023 from \tempurl}


\bibitem[Li and Liu(2021)]%
        {review_study_of_cyber-attacks}
\bibfield{author}{\bibinfo{person}{Yuchong Li} {and} \bibinfo{person}{Qinghui Liu}.} \bibinfo{year}{2021}\natexlab{}.
\newblock \showarticletitle{A comprehensive review study of cyber-attacks and cyber security; Emerging trends and recent developments}.
\newblock \bibinfo{journal}{\emph{Energy Reports}}  \bibinfo{volume}{7} (\bibinfo{year}{2021}), \bibinfo{pages}{8176--8186}.
\newblock
\showISSN{2352-4847}
\urldef\tempurl%
\url{https://doi.org/10.1016/j.egyr.2021.08.126}
\showDOI{\tempurl}


\bibitem[Lockfale(2023)]%
        {osint}
\bibfield{author}{\bibinfo{person}{Lockfale}.} \bibinfo{year}{2023}\natexlab{}.
\newblock \bibinfo{booktitle}{\emph{OSINT-Framework}}.
\newblock
\urldef\tempurl%
\url{https://github.com/lockfale/OSINT-Framework}
\showURL{%
Retrieved December 20, 2023 from \tempurl}


\bibitem[Lyon(2009)]%
        {nmap}
\bibfield{author}{\bibinfo{person}{Gordon~Fyodor Lyon}.} \bibinfo{year}{2009}\natexlab{}.
\newblock \bibinfo{booktitle}{\emph{Nmap Network Scanning: The Official Nmap Project Guide to Network Discovery and Security Scanning}}.
\newblock \bibinfo{publisher}{Insecure}.
\newblock
\showISBNx{0979958717}


\bibitem[Malik and Changalvala(2019)]%
        {Fighting_AI_with_AI}
\bibfield{author}{\bibinfo{person}{Hafiz Malik} {and} \bibinfo{person}{Raghavendar Changalvala}.} \bibinfo{year}{2019}\natexlab{}.
\newblock \showarticletitle{Fighting AI with AI: Fake Speech Detection Using Deep Learning}.
\newblock \bibinfo{journal}{\emph{Audio Engineering Society Conference: 2019 AES International Conference on Audio Forensics}} (\bibinfo{year}{2019}).
\newblock
\urldef\tempurl%
\url{http://www.aes.org/e-lib/browse.cfm?elib=20479}
\showURL{%
\tempurl}


\bibitem[McInerney(2023)]%
        {Hacking_AI_Steal_Models_from_MLflow_No_Exploit_Needed}
\bibfield{author}{\bibinfo{person}{Dan McInerney}.} \bibinfo{year}{2023}\natexlab{}.
\newblock \bibinfo{booktitle}{\emph{Hacking AI: Steal Models from MLflow, No Exploit Needed}}.
\newblock
\urldef\tempurl%
\url{https://protectai.com/blog/hacking-ai-steal-models-from-mlflow-no-exploit-needed}
\showURL{%
Retrieved November 9, 2023 from \tempurl}


\bibitem[Mehonic et~al\mbox{.}(2020)]%
        {memristor}
\bibfield{author}{\bibinfo{person}{Adnan Mehonic}, \bibinfo{person}{Abu Sebastian}, \bibinfo{person}{Bipin Rajendran}, \bibinfo{person}{Osvaldo Simeone}, \bibinfo{person}{Eleni Vasilaki}, {and} \bibinfo{person}{Anthony~J. Kenyon}.} \bibinfo{year}{2020}\natexlab{}.
\newblock \showarticletitle{Memristors—From In-Memory Computing, Deep Learning Acceleration, and Spiking Neural Networks to the Future of Neuromorphic and Bio-Inspired Computing}.
\newblock \bibinfo{journal}{\emph{Advanced Intelligent Systems}} \bibinfo{volume}{2}, \bibinfo{number}{11} (\bibinfo{year}{2020}).
\newblock
\urldef\tempurl%
\url{https://doi.org/10.1002/aisy.202000085}
\showDOI{\tempurl}


\bibitem[Mehrabi et~al\mbox{.}(2022)]%
        {Survey_on_Bias_and_Fairness_in_Machine_Learning}
\bibfield{author}{\bibinfo{person}{Ninareh Mehrabi}, \bibinfo{person}{Fred Morstatter}, \bibinfo{person}{Nripsuta Saxena}, \bibinfo{person}{Kristina Lerman}, {and} \bibinfo{person}{Aram Galstyan}.} \bibinfo{year}{2022}\natexlab{}.
\newblock \bibinfo{title}{A Survey on Bias and Fairness in Machine Learning}.
\newblock
\newblock
\showeprint[arxiv]{1908.09635}~[cs.LG]


\bibitem[Miessler(2023)]%
        {The_AI_Attack_Surface_Map_v1.0}
\bibfield{author}{\bibinfo{person}{Daniel Miessler}.} \bibinfo{year}{2023}\natexlab{}.
\newblock \bibinfo{booktitle}{\emph{The AI Attack Surface Map v1.0}}.
\newblock
\urldef\tempurl%
\url{https://danielmiessler.com/p/the-ai-attack-surface-map-v1-0/}
\showURL{%
Retrieved November 10, 2023 from \tempurl}


\bibitem[Mirsky(2022)]%
        {Discussion_Paper_The_Integrity_of_Medical_AI}
\bibfield{author}{\bibinfo{person}{Yisroel Mirsky}.} \bibinfo{year}{2022}\natexlab{}.
\newblock \showarticletitle{Discussion Paper: The Integrity of Medical AI}.
\newblock \bibinfo{journal}{\emph{Proceedings of the 1st Workshop on Security Implications of Deepfakes and Cheapfakes, Association for Computing Machinery}} (\bibinfo{year}{2022}), \bibinfo{pages}{31–33}.
\newblock
\showISBNx{9781450391788}
\urldef\tempurl%
\url{https://doi.org/10.1145/3494109.3527191}
\showDOI{\tempurl}


\bibitem[Morris et~al\mbox{.}(2020)]%
        {TextAttack}
\bibfield{author}{\bibinfo{person}{John Morris}, \bibinfo{person}{Eli Lifland}, \bibinfo{person}{Jin~Yong Yoo}, \bibinfo{person}{Jake Grigsby}, \bibinfo{person}{Di Jin}, {and} \bibinfo{person}{Yanjun Qi}.} \bibinfo{year}{2020}\natexlab{}.
\newblock \showarticletitle{TextAttack: A Framework for Adversarial Attacks, Data Augmentation, and Adversarial Training in NLP}.
\newblock \bibinfo{journal}{\emph{Proceedings of the 2020 Conference on Empirical Methods in Natural Language Processing: System Demonstrations}} (\bibinfo{year}{2020}), \bibinfo{pages}{119--126}.
\newblock


\bibitem[Moshe et~al\mbox{.}(2022)]%
        {The_Security_of_Deep_Learning_Defences_for_Medical_Imaging}
\bibfield{author}{\bibinfo{person}{Levy Moshe}, \bibinfo{person}{Amit Guy}, \bibinfo{person}{Elovici Yuval}, {and} \bibinfo{person}{Mirsky Yisroel}.} \bibinfo{year}{2022}\natexlab{}.
\newblock \bibinfo{title}{The Security of Deep Learning Defences for Medical Imaging}.
\newblock
\newblock
\showeprint[arxiv]{2201.08661}~[cs.CR]


\bibitem[Mozes et~al\mbox{.}(2023)]%
        {Use_of_LLMs_for_Illicit_Purposes}
\bibfield{author}{\bibinfo{person}{Maximilian Mozes}, \bibinfo{person}{Xuanli He}, \bibinfo{person}{Bennett Kleinberg}, {and} \bibinfo{person}{Lewis~D. Griffin}.} \bibinfo{year}{2023}\natexlab{}.
\newblock \bibinfo{title}{Use of LLMs for Illicit Purposes: Threats, Prevention Measures, and Vulnerabilities}.
\newblock
\newblock
\showeprint[arxiv]{2308.12833}~[cs.CL]


\bibitem[Ossman and Lehto(2023)]%
        {pyobfuscate}
\bibfield{author}{\bibinfo{person}{Pierre Ossman} {and} \bibinfo{person}{Niko~and Lehto}.} \bibinfo{year}{2023}\natexlab{}.
\newblock \bibinfo{booktitle}{\emph{pyobfuscate}}.
\newblock
\urldef\tempurl%
\url{https://github.com/astrand/pyobfuscate}
\showURL{%
Retrieved November 10, 2023 from \tempurl}


\bibitem[Papakipos and Bitton(2022)]%
        {AugLy}
\bibfield{author}{\bibinfo{person}{Zoe Papakipos} {and} \bibinfo{person}{Joanna Bitton}.} \bibinfo{year}{2022}\natexlab{}.
\newblock \bibinfo{title}{AugLy: Data Augmentations for Robustness}.
\newblock
\newblock
\showeprint[arxiv]{2201.06494}~[cs.AI]


\bibitem[Papernot et~al\mbox{.}(2016)]%
        {papernot2016distillation}
\bibfield{author}{\bibinfo{person}{Nicolas Papernot}, \bibinfo{person}{Patrick McDaniel}, \bibinfo{person}{Xi Wu}, \bibinfo{person}{Somesh Jha}, {and} \bibinfo{person}{Ananthram Swami}.} \bibinfo{year}{2016}\natexlab{}.
\newblock \bibinfo{title}{Distillation as a Defense to Adversarial Perturbations against Deep Neural Networks}.
\newblock
\newblock
\showeprint[arxiv]{1511.04508}~[cs.CR]


\bibitem[Park et~al\mbox{.}(2022)]%
        {fugio}
\bibfield{author}{\bibinfo{person}{Sunnyeo Park}, \bibinfo{person}{Daejun Kim}, \bibinfo{person}{Suman Jana}, {and} \bibinfo{person}{Sooel Son}.} \bibinfo{year}{2022}\natexlab{}.
\newblock \showarticletitle{{FUGIO}: Automatic Exploit Generation for {PHP} Object Injection Vulnerabilities}.
\newblock \bibinfo{journal}{\emph{31st USENIX Security Symposium (USENIX Security 22)}} (\bibinfo{year}{2022}), \bibinfo{pages}{197--214}.
\newblock
\showISBNx{978-1-939133-31-1}
\urldef\tempurl%
\url{https://www.usenix.org/conference/usenixsecurity22/presentation/park-sunnyeo}
\showURL{%
\tempurl}


\bibitem[Patel(2020)]%
        {How_AI_is_already_being_poisoned_against_you}
\bibfield{author}{\bibinfo{person}{Andrew Patel}.} \bibinfo{year}{2020}\natexlab{}.
\newblock \bibinfo{booktitle}{\emph{How AI is already being poisoned against you}}.
\newblock
\urldef\tempurl%
\url{https://blog.f-secure.com/how-ai-is-already-being-poisoned-against-you/}
\showURL{%
Retrieved August 20, 2023 from \tempurl}


\bibitem[Paul(2023)]%
        {Cybersec-stats-2023}
\bibfield{author}{\bibinfo{person}{Shibu Paul}.} \bibinfo{year}{2023}\natexlab{}.
\newblock \bibinfo{booktitle}{\emph{Cybersecurity Statistics and Predictions for 2023}}.
\newblock
\urldef\tempurl%
\url{https://cxotoday.com/specials/cybersecurity-statistics-and-predictions-for-2023/}
\showURL{%
Retrieved November 23, 2023 from \tempurl}


\bibitem[Pegoraro et~al\mbox{.}(2023)]%
        {To_ChatGPT_or_not_to_ChatGPT_That_is_the_question}
\bibfield{author}{\bibinfo{person}{Alessandro Pegoraro}, \bibinfo{person}{Kavita Kumari}, \bibinfo{person}{Hossein Fereidooni}, {and} \bibinfo{person}{Ahmad-Reza Sadeghi}.} \bibinfo{year}{2023}\natexlab{}.
\newblock \bibinfo{title}{To ChatGPT, or not to ChatGPT: That is the question!}
\newblock
\newblock
\showeprint[arxiv]{2304.01487}~[cs.LG]


\bibitem[Perry et~al\mbox{.}(2022)]%
        {Do_Users_Write_More_Insecure_Code_with_AI_Assistants}
\bibfield{author}{\bibinfo{person}{Neil Perry}, \bibinfo{person}{Megha Srivastava}, \bibinfo{person}{Deepak Kumar}, {and} \bibinfo{person}{Dan Boneh}.} \bibinfo{year}{2022}\natexlab{}.
\newblock \bibinfo{title}{Do Users Write More Insecure Code with AI Assistants?}
\newblock
\newblock
\showeprint[arxiv]{2211.03622}~[cs.CR]


\bibitem[Potti(2023)]%
        {security_ai_workbench}
\bibfield{author}{\bibinfo{person}{Sunil Potti}.} \bibinfo{year}{2023}\natexlab{}.
\newblock \bibinfo{booktitle}{\emph{Supercharging security with generative AI}}.
\newblock
\urldef\tempurl%
\url{https://cloud.google.com/blog/products/identity-security/rsa-google-cloud-security-ai-workbench-generative-ai}
\showURL{%
Retrieved December 19, 2023 from \tempurl}


\bibitem[protect ai(2023)]%
        {dlflow}
\bibfield{author}{\bibinfo{person}{protect ai}.} \bibinfo{year}{2023}\natexlab{}.
\newblock \bibinfo{booktitle}{\emph{DLflow}}.
\newblock
\urldef\tempurl%
\url{https://github.com/protectai/Snaike-MLflow/tree/main/DLflow}
\showURL{%
Retrieved December 9, 2023 from \tempurl}


\bibitem[Qi et~al\mbox{.}(2023)]%
        {ft-llms-vulnerable}
\bibfield{author}{\bibinfo{person}{Xiangyu Qi}, \bibinfo{person}{Yi Zeng}, \bibinfo{person}{Tinghao Xie}, \bibinfo{person}{Pin-Yu Chen}, \bibinfo{person}{Ruoxi Jia}, \bibinfo{person}{Prateek Mittal}, {and} \bibinfo{person}{Peter Henderson}.} \bibinfo{year}{2023}\natexlab{}.
\newblock \bibinfo{title}{Fine-tuning Aligned Language Models Compromises Safety, Even When Users Do Not Intend To!}
\newblock
\newblock
\showeprint[arxiv]{2310.03693}~[cs.CL]


\bibitem[Quintero-Bonilla and Martín~del Rey(2020)]%
        {apt-lateral}
\bibfield{author}{\bibinfo{person}{Santiago Quintero-Bonilla} {and} \bibinfo{person}{Angel Martín~del Rey}.} \bibinfo{year}{2020}\natexlab{}.
\newblock \showarticletitle{A New Proposal on the Advanced Persistent Threat: A Survey}.
\newblock \bibinfo{journal}{\emph{Applied Sciences}} \bibinfo{volume}{10}, \bibinfo{number}{11} (\bibinfo{year}{2020}).
\newblock
\showISSN{2076-3417}
\urldef\tempurl%
\url{https://doi.org/10.3390/app10113874}
\showDOI{\tempurl}


\bibitem[Reddy et~al\mbox{.}(2022)]%
        {Data_Breaches_in_Healthcare_Security_Systems}
\bibfield{author}{\bibinfo{person}{Jahnavi Reddy}, \bibinfo{person}{Nelly Elsayed}, \bibinfo{person}{Zag ElSayed}, {and} \bibinfo{person}{Murat Ozer}.} \bibinfo{year}{2022}\natexlab{}.
\newblock \bibinfo{title}{Data Breaches in Healthcare Security Systems}.
\newblock
\newblock
\showeprint[arxiv]{2111.00582}~[cs.CR]


\bibitem[Roy et~al\mbox{.}(2023)]%
        {MITRE}
\bibfield{author}{\bibinfo{person}{Shanto Roy}, \bibinfo{person}{Emmanouil Panaousis}, \bibinfo{person}{Cameron Noakes}, \bibinfo{person}{Aron Laszka}, \bibinfo{person}{Sakshyam Panda}, {and} \bibinfo{person}{George Loukas}.} \bibinfo{year}{2023}\natexlab{}.
\newblock \bibinfo{title}{SoK: The MITRE ATT\&CK Framework in Research and Practice}.
\newblock
\newblock
\showeprint[arxiv]{2304.07411}~[cs.CR]


\bibitem[Schade(2023)]%
        {How_your_data_is_used_to_improve_model_performance}
\bibfield{author}{\bibinfo{person}{Michael Schade}.} \bibinfo{year}{2023}\natexlab{}.
\newblock \bibinfo{booktitle}{\emph{How your data is used to improve model performance}}.
\newblock
\urldef\tempurl%
\url{https://help.openai.com/en/articles/5722486-how-your-data-is-used-to-improve-model-performance}
\showURL{%
Retrieved October 8, 2023 from \tempurl}


\bibitem[Shanahan(2023)]%
        {anthropomorphism}
\bibfield{author}{\bibinfo{person}{Murray Shanahan}.} \bibinfo{year}{2023}\natexlab{}.
\newblock \bibinfo{title}{Talking About Large Language Models}.
\newblock
\newblock
\showeprint[arxiv]{2212.03551}~[cs.CL]


\bibitem[Shankar and Kumar(2021)]%
        {Counterfit}
\bibfield{author}{\bibinfo{person}{Ram Shankar} {and} \bibinfo{person}{Siva Kumar}.} \bibinfo{year}{2021}\natexlab{}.
\newblock \bibinfo{booktitle}{\emph{AI security risk assessment using Counterfit}}.
\newblock
\urldef\tempurl%
\url{https://www.microsoft.com/en-us/security/blog/2021/05/03/ai-security-risk-assessment-using-counterfit}
\showURL{%
Retrieved October 12, 2023 from \tempurl}


\bibitem[Shen et~al\mbox{.}(2021)]%
        {Model_Stealing_Attacks_Against_Inductive_Graph_Neural_Networks}
\bibfield{author}{\bibinfo{person}{Yun Shen}, \bibinfo{person}{Xinlei He}, \bibinfo{person}{Yufei Han}, {and} \bibinfo{person}{Yang Zhang}.} \bibinfo{year}{2021}\natexlab{}.
\newblock \bibinfo{title}{Model Stealing Attacks Against Inductive Graph Neural Networks}.
\newblock
\newblock
\showeprint[arxiv]{2112.08331}~[cs.CR]


\bibitem[Sims(2023)]%
        {Eyespy_Proof_Of_Concept}
\bibfield{author}{\bibinfo{person}{Jeff Sims}.} \bibinfo{year}{2023}\natexlab{}.
\newblock \bibinfo{booktitle}{\emph{Eyespy Proof-Of-Concept}}.
\newblock
\urldef\tempurl%
\url{https://www.hyas.com/blog/eyespy-proof-of-concept}
\showURL{%
Retrieved June 9, 2023 from \tempurl}


\bibitem[Spencer(2022)]%
        {credential-theft-report}
\bibfield{author}{\bibinfo{person}{Patrick Spencer}.} \bibinfo{year}{2022}\natexlab{}.
\newblock \bibinfo{booktitle}{\emph{Prime Cyber Targets According to the 2022 Verizon DBIR}}.
\newblock
\urldef\tempurl%
\url{https://www.kiteworks.com/third-party-risk/prime-cyber-targets-according-to-the-2022-verizon-dbir/}
\showURL{%
Retrieved November 11, 2023 from \tempurl}


\bibitem[Spring et~al\mbox{.}(2020)]%
        {On_managing_vulnerabilities_in_AI_ML_systems}
\bibfield{author}{\bibinfo{person}{Jonathan~M. Spring}, \bibinfo{person}{April Galyardt}, \bibinfo{person}{Allen~D. Householder}, {and} \bibinfo{person}{Nathan VanHoudnos}.} \bibinfo{year}{2020}\natexlab{}.
\newblock \showarticletitle{On managing vulnerabilities in AI/ML systems}.
\newblock \bibinfo{journal}{\emph{New Security Paradigms Workshop 2020}} (\bibinfo{year}{2020}).
\newblock
\urldef\tempurl%
\url{https://doi.org/10.1145/3442167.3442177}
\showDOI{\tempurl}


\bibitem[Szegedy et~al\mbox{.}(2014)]%
        {Intriguing_properties_of_neural_networks}
\bibfield{author}{\bibinfo{person}{Christian Szegedy}, \bibinfo{person}{Wojciech Zaremba}, \bibinfo{person}{Ilya Sutskever}, \bibinfo{person}{Joan Bruna}, \bibinfo{person}{Dumitru Erhan}, \bibinfo{person}{Ian Goodfellow}, {and} \bibinfo{person}{Rob Fergus}.} \bibinfo{year}{2014}\natexlab{}.
\newblock \bibinfo{title}{Intriguing properties of neural networks}.
\newblock
\newblock
\showeprint[arxiv]{1312.6199}~[cs.CV]


\bibitem[Team(2023)]%
        {Unravelling_the_Attack_Surface_of_AI_Systems}
\bibfield{author}{\bibinfo{person}{Counter Threat Unit~Research Team}.} \bibinfo{year}{2023}\natexlab{}.
\newblock \bibinfo{booktitle}{\emph{Unravelling the Attack Surface of AI Systems}}.
\newblock
\urldef\tempurl%
\url{https://www.secureworks.com/blog/unravelling-the-attack-surface-of-ai-systems}
\showURL{%
Retrieved September 29, 2023 from \tempurl}


\bibitem[Truong et~al\mbox{.}(2020)]%
        {Artificial_Intelligence_in_the_Cyber_Domain_Offense_and_Defense}
\bibfield{author}{\bibinfo{person}{Thanh~Cong Truong}, \bibinfo{person}{Quoc~Bao Diep}, {and} \bibinfo{person}{Ivan Zelinka}.} \bibinfo{year}{2020}\natexlab{}.
\newblock \showarticletitle{Artificial Intelligence in the Cyber Domain: Offense and Defense}.
\newblock \bibinfo{journal}{\emph{Symmetry}} \bibinfo{volume}{12}, \bibinfo{number}{3} (\bibinfo{year}{2020}), \bibinfo{pages}{410}.
\newblock
\showISSN{2073-8994}
\urldef\tempurl%
\url{https://doi.org/10.3390/sym12030410}
\showDOI{\tempurl}


\bibitem[Vinayakumar et~al\mbox{.}(2019)]%
        {Robust_Intelligent_Malware_Detection_Using_Deep_Learning}
\bibfield{author}{\bibinfo{person}{R. Vinayakumar}, \bibinfo{person}{Mamoun Alazab}, \bibinfo{person}{K.~P. Soman}, \bibinfo{person}{Prabaharan Poornachandran}, {and} \bibinfo{person}{Sitalakshmi Venkatraman}.} \bibinfo{year}{2019}\natexlab{}.
\newblock \showarticletitle{Robust Intelligent Malware Detection Using Deep Learning}.
\newblock \bibinfo{journal}{\emph{IEEE Access}}  \bibinfo{volume}{7} (\bibinfo{year}{2019}), \bibinfo{pages}{46717--46738}.
\newblock
\urldef\tempurl%
\url{https://doi.org/10.1109/ACCESS.2019.2906934}
\showDOI{\tempurl}


\bibitem[Vincent et~al\mbox{.}(2015)]%
        {trojan-manfac}
\bibfield{author}{\bibinfo{person}{Hannah Vincent}, \bibinfo{person}{Lee Wells}, \bibinfo{person}{Pablo Tarazaga}, {and} \bibinfo{person}{Jaime Camelio}.} \bibinfo{year}{2015}\natexlab{}.
\newblock \showarticletitle{Trojan Detection and Side-channel Analyses for Cyber-security in Cyber-physical Manufacturing Systems}.
\newblock \bibinfo{journal}{\emph{43rd North American Manufacturing Research Conference, Procedia Manufacturing}}  \bibinfo{volume}{1} (\bibinfo{year}{2015}), \bibinfo{pages}{77--85}.
\newblock
\showISSN{2351-9789}
\urldef\tempurl%
\url{https://doi.org/10.1016/j.promfg.2015.09.065}
\showDOI{\tempurl}


\bibitem[Vincent(2016)]%
        {Twitter_taught_Microsofts_AI_chatbot}
\bibfield{author}{\bibinfo{person}{James Vincent}.} \bibinfo{year}{2016}\natexlab{}.
\newblock \bibinfo{booktitle}{\emph{Twitter taught Microsoft’s AI chatbot to be a racist asshole in less than a day}}.
\newblock
\urldef\tempurl%
\url{https://www.theverge.com/2016/3/24/11297050/tay-microsoft-chatbot-racist}
\showURL{%
Retrieved September 20, 2023 from \tempurl}


\bibitem[von Platen et~al\mbox{.}(2022)]%
        {diffusers}
\bibfield{author}{\bibinfo{person}{Patrick von Platen}, \bibinfo{person}{Suraj Patil}, \bibinfo{person}{Anton Lozhkov}, \bibinfo{person}{Pedro Cuenca}, \bibinfo{person}{Nathan Lambert}, \bibinfo{person}{Kashif Rasul}, \bibinfo{person}{Mishig Davaadorj}, {and} \bibinfo{person}{Thomas Wolf}.} \bibinfo{year}{2022}\natexlab{}.
\newblock \bibinfo{booktitle}{\emph{Diffusers: State-of-the-art diffusion models}}.
\newblock
\urldef\tempurl%
\url{https://github.com/huggingface/diffusers}
\showURL{%
Retrieved August 19, 2023 from \tempurl}


\bibitem[Wickens(2023)]%
        {Insane_in_The_Supply_Chain}
\bibfield{author}{\bibinfo{person}{Eoin Wickens}.} \bibinfo{year}{2023}\natexlab{}.
\newblock \bibinfo{booktitle}{\emph{Insane in The Supply Chain}}.
\newblock
\urldef\tempurl%
\url{https://hiddenlayer.com/research/insane-in-the-supply-chain/}
\showURL{%
Retrieved December 9, 2023 from \tempurl}


\bibitem[Wickens et~al\mbox{.}(2022a)]%
        {The_Tactics_And_Techniques_of_Adversarial_Ml}
\bibfield{author}{\bibinfo{person}{Eoin Wickens}, \bibinfo{person}{Marta Janus}, {and} \bibinfo{person}{Tom Bonner}.} \bibinfo{year}{2022}\natexlab{a}.
\newblock \bibinfo{booktitle}{\emph{The Tactics And Techniques of Adversarial Ml}}.
\newblock
\urldef\tempurl%
\url{https://hiddenlayer.com/research/the-tactics-and-techniques-of-adversarial-ml/}
\showURL{%
Retrieved September 21, 2023 from \tempurl}


\bibitem[Wickens et~al\mbox{.}(2022b)]%
        {Weaponizing_Machine_Learning_Models_With_Ransomware}
\bibfield{author}{\bibinfo{person}{Eoin Wickens}, \bibinfo{person}{Marta Janus}, {and} \bibinfo{person}{Tom Bonner}.} \bibinfo{year}{2022}\natexlab{b}.
\newblock \bibinfo{booktitle}{\emph{Weaponizing Machine Learning Models With Ransomware}}.
\newblock
\urldef\tempurl%
\url{https://hiddenlayer.com/research/weaponizing-machine-learning-models-with-ransomware/}
\showURL{%
Retrieved September 19, 2023 from \tempurl}


\bibitem[Wickens et~al\mbox{.}(2023)]%
        {Crossing_the_Rubika___The_Use_and_Abuse_of_AI_Cloud_Services}
\bibfield{author}{\bibinfo{person}{Eoin Wickens}, \bibinfo{person}{Marta Janus}, {and} \bibinfo{person}{Tom Bonner}.} \bibinfo{year}{2023}\natexlab{}.
\newblock \bibinfo{booktitle}{\emph{Crossing the Rubika - The Use and Abuse of AI Cloud Services}}.
\newblock
\urldef\tempurl%
\url{https://hiddenlayer.com/research/crossing-the-rubika-the-use-and-abuse-of-ai-cloud-services/}
\showURL{%
Retrieved July 29, 2023 from \tempurl}


\bibitem[Wu et~al\mbox{.}(2022)]%
        {Training_Time_Attack_for_Cooperative_MultiAgent_RL}
\bibfield{author}{\bibinfo{person}{Siyang Wu}, \bibinfo{person}{Tonghan Wang}, \bibinfo{person}{Xiaoran Wu}, \bibinfo{person}{Jingfeng Zhang}, \bibinfo{person}{Yujing Hu}, \bibinfo{person}{Changjie Fan}, {and} \bibinfo{person}{Chongjie Zhang}.} \bibinfo{year}{2022}\natexlab{}.
\newblock \showarticletitle{Model and Method: Training-Time Attack for Cooperative Multi-Agent Reinforcement Learning}.
\newblock \bibinfo{journal}{\emph{Deep Reinforcement Learning Workshop NeurIPS 2022}} (\bibinfo{year}{2022}).
\newblock
\urldef\tempurl%
\url{https://openreview.net/forum?id=cZSNk8veQW7}
\showURL{%
\tempurl}


\bibitem[Wu et~al\mbox{.}(2023)]%
        {wu2023intelldragonfly}
\bibfield{author}{\bibinfo{person}{Xingchen Wu}, \bibinfo{person}{Qin Qiu}, \bibinfo{person}{Jiaqi Li}, {and} \bibinfo{person}{Yang Zhao}.} \bibinfo{year}{2023}\natexlab{}.
\newblock \bibinfo{title}{Intell-dragonfly: A Cybersecurity Attack Surface Generation Engine Based On Artificial Intelligence-generated Content Technology}.
\newblock
\newblock
\showeprint[arxiv]{2311.00240}~[cs.CR]


\bibitem[Xu et~al\mbox{.}(2016)]%
        {Automatically_Evading_Classifiers_A_Case_Study_on_PDF_Malware_Classifiers}
\bibfield{author}{\bibinfo{person}{Weilin Xu}, \bibinfo{person}{Yanjun Qi}, {and} \bibinfo{person}{David Evans}.} \bibinfo{year}{2016}\natexlab{}.
\newblock \showarticletitle{Automatically Evading Classifiers: A Case Study on PDF Malware Classifiers}.
\newblock \bibinfo{journal}{\emph{Network and Distributed System Security Symposium}} (\bibinfo{year}{2016}).
\newblock
\urldef\tempurl%
\url{https://api.semanticscholar.org/CorpusID:787013}
\showURL{%
\tempurl}


\bibitem[Young and Ewing(2022)]%
        {What_are_cloud_services}
\bibfield{author}{\bibinfo{person}{Bob Young} {and} \bibinfo{person}{Marc Ewing}.} \bibinfo{year}{2022}\natexlab{}.
\newblock \bibinfo{booktitle}{\emph{What are cloud services}}.
\newblock
\urldef\tempurl%
\url{https://www.redhat.com/en/topics/cloud-computing/what-are-cloud-services}
\showURL{%
Retrieved November 2, 2023 from \tempurl}


\bibitem[Zaharia et~al\mbox{.}(2018)]%
        {mlflow}
\bibfield{author}{\bibinfo{person}{Matei~A. Zaharia}, \bibinfo{person}{Andrew Chen}, \bibinfo{person}{Aaron Davidson}, \bibinfo{person}{Ali Ghodsi}, \bibinfo{person}{Sue~Ann Hong}, \bibinfo{person}{Andy Konwinski}, \bibinfo{person}{Siddharth Murching}, \bibinfo{person}{Tomas Nykodym}, \bibinfo{person}{Paul Ogilvie}, \bibinfo{person}{Mani Parkhe}, \bibinfo{person}{Fen Xie}, {and} \bibinfo{person}{Corey Zumar}.} \bibinfo{year}{2018}\natexlab{}.
\newblock \showarticletitle{Accelerating the Machine Learning Lifecycle with MLflow}.
\newblock \bibinfo{journal}{\emph{IEEE Data Eng. Bull.}}  \bibinfo{volume}{41} (\bibinfo{year}{2018}), \bibinfo{pages}{39--45}.
\newblock
\urldef\tempurl%
\url{https://api.semanticscholar.org/CorpusID:83459546}
\showURL{%
\tempurl}


\bibitem[Zerofox(2016)]%
        {SNAP_R}
\bibfield{author}{\bibinfo{person}{Zerofox}.} \bibinfo{year}{2016}\natexlab{}.
\newblock \bibinfo{booktitle}{\emph{SNAP\_R: A {M}achine {L}earning based social media pen-testing tool}}.
\newblock
\urldef\tempurl%
\url{https://github.com/zerofox-oss/SNAP_R}
\showURL{%
Retrieved September 28, 2023 from \tempurl}


\bibitem[Zetter(2019)]%
        {Hospital_viruses_Fake_cancerous_nodes_in_CT_scans_created_by_malware_trick_radiologists}
\bibfield{author}{\bibinfo{person}{Kim Zetter}.} \bibinfo{year}{2019}\natexlab{}.
\newblock \bibinfo{booktitle}{\emph{Hospital viruses: Fake cancerous nodes in CT scans, created by malware, trick radiologists}}.
\newblock
\urldef\tempurl%
\url{https://www.washingtonpost.com/technology/2019/04/03/hospital-viruses-fake-cancerous-nodes-ct-scans-created-by-malware-trick-radiologists/}
\showURL{%
Retrieved October 9, 2023 from \tempurl}


\bibitem[Zhang et~al\mbox{.}(2021)]%
        {multi-agent}
\bibfield{author}{\bibinfo{person}{Dan Zhang}, \bibinfo{person}{Gang Feng}, \bibinfo{person}{Yang Shi}, {and} \bibinfo{person}{Dipti Srinivasan}.} \bibinfo{year}{2021}\natexlab{}.
\newblock \showarticletitle{Physical Safety and Cyber Security Analysis of Multi-Agent Systems: A Survey of Recent Advances}.
\newblock \bibinfo{journal}{\emph{IEEE/CAA Journal of Automatica Sinica}} \bibinfo{volume}{8}, \bibinfo{number}{2} (\bibinfo{year}{2021}), \bibinfo{pages}{319--333}.
\newblock
\urldef\tempurl%
\url{https://doi.org/10.1109/JAS.2021.1003820}
\showDOI{\tempurl}


\bibitem[Zhao et~al\mbox{.}(2017)]%
        {Efficient_Label_Contamination_Attacks_Against_Black_Box_Learning_Models}
\bibfield{author}{\bibinfo{person}{Mengchen Zhao}, \bibinfo{person}{Bo An}, \bibinfo{person}{Wei Gao}, {and} \bibinfo{person}{Teng Zhang}.} \bibinfo{year}{2017}\natexlab{}.
\newblock \showarticletitle{Efficient Label Contamination Attacks Against Black-Box Learning Models}.
\newblock \bibinfo{journal}{\emph{Proceedings of the Twenty-Sixth International Joint Conference on Artificial Intelligence, {IJCAI-17}}} (\bibinfo{year}{2017}), \bibinfo{pages}{3945--3951}.
\newblock
\urldef\tempurl%
\url{https://doi.org/10.24963/ijcai.2017/551}
\showDOI{\tempurl}


\bibitem[Zou et~al\mbox{.}(2023)]%
        {Universal_and_Transferable_Adversarial_Attacks_on_Aligned_Language_Models}
\bibfield{author}{\bibinfo{person}{Andy Zou}, \bibinfo{person}{Zifan Wang}, \bibinfo{person}{J.~Zico Kolter}, {and} \bibinfo{person}{Matt Fredrikson}.} \bibinfo{year}{2023}\natexlab{}.
\newblock \bibinfo{title}{Universal and Transferable Adversarial Attacks on Aligned Language Models}.
\newblock
\newblock
\showeprint[arxiv]{2307.15043}~[cs.CL]


\end{thebibliography}

\appendix

\section{Basic Terms in Cybersecurity}


\begin{itemize}
    \item \textbf{Weakness}: An error, typically in the software code, that might lead to a vulnerability. It is a defined set, and hence does not expand.
    
    \item \textbf{Vulnerability}: A vulnerability in security refers to a weakness or opportunity in an information system that cybercriminals can exploit. Vulnerabilities are known weaknesses. it’s the pathway for malicious actors to access its target.
    
    \item \textbf{Exposure}: Incident in which the vulnerability has been taken advantage of by an unauthorised activity.
    
    \item \textbf{Exploit}: An exploit is the actual malicious code that cybercriminals use to take advantage of vulnerabilities and compromise the IT infrastructure. Exploits are how threats become attacks, and vulnerabilities are how exploits gain access to targeted systems.
    
    \item \textbf{Threat}: A threat is a potentially dangerous event that has not occurred but has the potential to cause damage if it does. Malware, ransomware, phishing, malicious code, and wrongfully accessing user login credentials are all examples of intentional threats.
    
    \item \textbf{Risks}: Risks are potentials for loss or damage when a threat exploits a vulnerability.
\end{itemize}
 
A common equation for calculating it is 
$$Risk = Threat \cdot Vulnerability \cdot Consequence$$


\section{Common Terms for Identifying Weaknesses}


\begin{itemize}
    \item \textbf{Common Vulnerabilities and Exposures (CVE)}: It is a compilation that comprises of publicly disclosed Vulnerabilities and Exposures. Each security flaw within this list is uniquely identified by a CVE ID, such as CVE-2019-1234567. These CVE IDs provide a dependable means for users to distinguish individual vulnerabilities and are assigned by a CVE Numbering Authority (CNA). Importantly, a CVE ID is allocated before the corresponding security advisory becomes publicly accessible, serving the purpose of maintaining the confidentiality of security flaws until an effective solution has been developed and tested. The CVE database logs real-world occurrences of vulnerabilities and exposures found in specific products.
    
    \item \textbf{Common Vulnerability Scoring System (CVSS)}: It is a way to evaluate the severity of a vulnerability. This is a set of open standards for assigning a number to a vulnerability to assess its severity. The scores range from 0.0 to 10.0, with higher numbers representing a higher degree of severity of the vulnerability.
    
    \item \textbf{Common Weaknesses Enumeration (CWE)}: Software weaknesses are often discussed and defined in the context of the CWEs, a community-developed list of common software security weaknesses in digital products. It does not refer to one particular example but provides definitions for widely seen defects. The more vulnerabilities in CVE, the better the definition of weaknesses in CWE, highlighting the importance of knowing the difference between the two sites.
    
\end{itemize}


\section{Abstraction levels}

In cybersecurity, abstraction levels refer to the hierarchical organisation of concepts and technologies. They range from high-level, language-agnostic principles to lower-level, implementation-specific details. They allow for a systematic understanding of security issues across different layers of complexity. We list out the common abstraction levels below

\begin{itemize}
    \item \textbf{Classes}: These are described in a highly abstract manner, independent of any particular programming language or technology. Weaknesses at this foundational level tend to be somewhat more specific, often providing details that encompass methods for both detection and prevention.
    \item \textbf{Variants}: Represent the most specific instances and are described with less detail.
    \item \textbf{Category}: An entry that encompasses a CWE list of other entries, all sharing a common characteristic viewpoint.
    \item \textbf{View}: A subset of CWE entries grouped together based on shared characteristics.
\end{itemize}

\section{MITRE ATT\&CK}



The MITRE ATT\&CK framework is designed for threat hunters, defenders, and red teams to classify attacks, identify attribution and objectives, and assess organisational risk. Organisations can use the framework to identify security gaps and prioritize mitigations based on risk. The 2022 version of ATT\&CK for Enterprise contains 14 Tactics, 193 Techniques, 401 Subtechniques, 135 Groups, 14 Campaigns, and 718 Pieces of Software. Table~\ref{table: MITRE Attack objectives table} lists all the tactics and attacker objectives within the framework.


\begin{table}[h]
\centering
    \begin{tabular}{|l|l|p{8cm}|}
    \hline
    \textbf{Sr. No.} & \textbf{Tactic} & \textbf{Attacker(s) Objective} \\
    \hline
    1. & Reconnaissance & Gather information for planning future operations \\
    2. & Resource Development & Establish resources to support operations \\
    3. & Initial Access & Gain entry into the network \\
    4. & Execution & Run malicious code \\
    5. & Persistence & Maintain foothold \\
    6. & Privilege Escalation & Gain higher-level permissions \\
    7. & Defense Evasion & Avoid detection \\
    8. & Credential Access & Steal account names and passwords \\
    9. & Discovery & Understand the environment \\
    10. & Lateral Movement & Move through the environment \\
    11. & Collection & Gather data of interest \\
    12. & Command and Control & Communicate with compromised systems \\
    13. & Exfiltration & Steal data \\
    14. & Impact & Manipulate, interrupt, or destroy systems and data \\
    \hline
    \end{tabular}
    \caption{All the 14 tactics in the MITRE ATT\&CK framework with their corresponding objectives.}
    \label{table: MITRE Attack objectives table}
\end{table}

\begin{itemize}
    \item \textbf{Tactics}: These represent the \emph{why} of an ATT\&CK technique or subtechnique. There are 14 tactics in the Enterprise ATT\&CK Matrix.
    \item \textbf{Techniques}: These represent \emph{how} an adversary achieves a tactical goal by performing an action. MITRE regularly updates the techniques discovered.
    \item \textbf{Subtechniques}: These are the more specific descriptions of adversarial behaviour used to achieve a goal. For example, an adversary may dump credentials by accessing the Local Security Authority (LSA) Secrets.
    \item \textbf{Procedures}: These are the specific implementations that adversaries use for techniques or subtechniques.
\end{itemize}

\end{document}